\documentclass[a4paper,11pt]{article}
\pdfoutput=1 
\usepackage{grffile}
\usepackage{aas_macros,braket}
\usepackage{jcappub,natbib} 
\usepackage[T1]{fontenc} 
\usepackage{color}
\usepackage{comment}

\usepackage{mathtools}
\usepackage{amsmath}
\allowdisplaybreaks
\usepackage{float}

\def\be{\begin{equation}}
\def\ee{\end{equation}}
\def\bea{\begin{eqnarray}}
\def\eea{\end{eqnarray}}

\def\ba#1\ea{\begin{align}#1\end{align}}

\def\apjs{Astrophys.\ J. Supp.}

\definecolor{green2}{cmyk}{1, 0, 1, 0.1}

\title{Tensor non-Gaussianity\\ in chiral scalar-tensor theories of gravity}

\author[a,b]{Nicola Bartolo,}
\author[c,d]{Luca Caloni,}
\author[a,b]{Giorgio Orlando,}
\author[b]{and Angelo Ricciardone}

\affiliation[a]{Dipartimento di Fisica e Astronomia ``G. Galilei'', Universit\`a degli Studi di Padova, via Marzolo 8, I-35131, Padova, Italy}
\affiliation[b]{INFN, Sezione di Padova, via Marzolo 8, I-35131, Padova, Italy}
\affiliation[c]{Dipartimento di Fisica e Scienze della Terra, Universit\`a degli Studi di Ferrara, via Saragat 1, I-44122 Ferrara, Italy}
\affiliation[d]{INFN, Sezione di Ferrara, via Saragat 1, I-44122 Ferrara, Italy}

\emailAdd{nicola.bartolo@pd.infn.it}
\emailAdd{luca.caloni@unife.it}
\emailAdd{giorgio.orlando@phd.unipd.it}
\emailAdd{angelo.ricciardone@pd.infn.it}

\abstract{Violation of parity symmetry in the gravitational sector, which manifests into unequal left and right circular polarization states of primordial gravitational waves, represents a way to test high-energy modifications to general relativity. In this paper we study inflation within recently proposed chiral scalar-tensor theories of gravity, that extend Chern-Simons gravity by including parity-violating operators containing first and second derivatives of the non-minimally coupled scalar (inflaton) field.
Given the degeneracy between different parity-violating theories at the level of the power spectrum statistics, we make a detailed analysis of the parity violation on primordial tensor non-Gaussianity. We show, with an explicit computation, that no new contributions arise in the graviton bispectra if the couplings in the new operators are constant in a pure de Sitter phase. On the other hand, if the coupling functions are time-dependent during inflation, the tensor bispectra acquire non-vanishing contributions from the parity-breaking operators even in the exact de Sitter limit, with maximal signal in the squeezed and equilateral configurations. We also comment on the consistency relation of the three-point function of tensor modes in this class of models and discuss prospects of detecting parity-breaking signatures through Cosmic Microwave Background $B$-mode bispectra.}

\begin{document}

\maketitle
\flushbottom

\section{Introduction}

Within the Einstein gravity framework, single (scalar) field slow-roll models of inflation \cite{Brout:1977ix,Sato:1980yn,Guth:1980zm,Starobinsky:1980te,Linde:1981mu,Albrecht:1982wi} are in accordance with measurements of the Cosmic Microwave Background (CMB), in particular with the recent data provided by WMAP and Planck missions~\cite{Hinshaw:2012aka,Aghanim:2018eyx,Akrami:2018odb,Akrami:2019izv}.
Due to the invariance under parity symmetry of general relativity, the two circular polarized right (R) and left (L)-handed polarization modes of primordial gravitational waves (PGWs) share exactly the same statistical properties and give the same contribution to the total tensor power spectrum. However, at very high energies, when inflation takes place, it might be that signatures of modification to Einstein gravity are left imprinted on the inflationary quantum fluctuations. In particular, several candidates of quantum gravity admit the presence of additional parity breaking gravitational terms yielding to the violation of parity in the gravitational interaction. 

One  example of these terms is the so-called gravitational four-dimensional Chern-Simons term coupled to a scalar field, proposed for the first time in \cite{Lue:1998mq}, and which naturally appears in the context of anomaly cancellation in string theory (via the so-called Green-Schwarz mechanism \cite{Green:1984sg,Polchinski:1998rr,Choi:1999zy,Alexander:2004us,Lyth:2005jf,Alexander:2009tp,Kamada:2020jaf}) and in loop quantum gravity \cite{Ashtekar:1989,Taveras:2008yf,Calcagni:2009xz,Gates:2009pt,Mercuri:2009zt}. The four-dimensional Chern-Simons operator is also commonly introduced as a low energy effective field theory in an expansion in the curvature invariants \cite{Weinberg:2008hq}. Indeed, it represents the fully covariant operator that breaks parity with the lowest number of derivatives.
Another example is the so-called three-dimensional Chern-Simons term which arises from  Horava-Lifshitz gravity \cite{Horava:2009uw,Wang:2017brl}. 

In slow-roll inflation with the additional presence of these terms, the parity violation generates a different behavior in the propagation of the R and L-handed polarization modes of PGWs. At linear level this parity breaking is quantified by the relative difference between the super-horizon R and L-handed tensor power spectra, which in the literature is usually referred to chirality of PGWs. However, both three \cite{Takahashi:2009wc,Wang:2012fi} and four \cite{Alexander:2004wk,Satoh:2010ep,Myung:2014jha,Alexander:2016hxk,Mylova:2019jrj} dimensional Chern-Simons terms are expected to give a low level of chirality assuming  adiabaticity for PGWs. In fact, the natural production of a large level of chirality in the tensor power spectrum is typical of other scenarios, such as the so-called Chromonatural inflation scenario~\cite{Maleknejad:2011jw,Dimastrogiovanni:2012ew,Adshead:2013nka,Adshead:2013qp,Maleknejad:2016qjz,Obata:2016xcr,Obata:2016tmo,Dimastrogiovanni:2016fuu,Domcke:2018rvv,Domcke:2018gfr,Maleknejad:2018nxz,Papageorgiou:2018rfx,McDonough:2018xzh,Papageorgiou:2019ecb,Mirzagholi:2020irt}, where an additional SU(2) gauge boson is coupled to a pseudo-scalar (axion-like) field through a Chern-Simons like operator. 

Recently, based on the Chern-Simons four-dimensional term, another ghost-free parity-breaking theory of gravity has been proposed in \cite{Crisostomi:2017ugk} by including terms with higher order derivatives of the non-minimally coupled scalar field. The level of parity breaking induced by these theories in the primordial tensor power spectrum has been explored in \cite{Qiao:2019hkz}. With respect to the four dimensional Chern-Simons scenario one of the distinguishable features of higher derivatives of the coupling scalar field is that they lead to the velocity birefringence phenomenon, i.e. they induce a difference in the propagation speed of the two circular polarizations. However, despite this distinct feature, the final amount of circular polarization (chirality) produced at the end of inflation is degenerate with respect to e.g. four-dimensional Chern-Simons gravity, yielding a small level of chirality in the tensor power spectrum. The limited amount of parity breaking operators quadratic in PGWs \cite{Weinberg:2008hq,Creminelli:2014wna,Baumann:2015xxa} is also responsible for such a degeneracy. 

From the observational point of view, it is well known that parity breaking signatures in the primordial power spectra leave distinct imprints in the CMB $TB$ and $EB$ angular correlators (see, e.g., the original Ref. \cite{Lue:1998mq}). However, it has been recently shown that CMB angular power spectra, even in the ideal case, are able to probe parity breaking in the primordial Universe only for models with maximum chirality \cite{Gerbino:2016mqb}. 

Parity breaking from the primordial Universe can be detected also by ground and space-based interferometers \cite{Seto:2008sr,Smith:2016jqs,Thorne:2017jft,Domcke:2019zls,Seto:2020zxw,Orlando:2020oko}. In fact, even if in principle the planar geometry of the planned space-based interferometers makes them unable to detect such signatures, the possibility of cross-correlating signals from different detectors and the dipolar anisotropy induced by the motion of the solar system with respect to the cosmic reference frame, allow for a possible measurement. Given the current constraints on the tensor power spectrum from CMB experiments, the forthcoming interferometers (like, e.g., LISA or Einstein Telescope experiments \cite{Bartolo:2016ami, Maggiore:2019uih}) are expected to be able to detect the primordial background of gravitational waves only for a subset of inflationary scenarios where PGWs have a blue tensor tilt, i.e. under the condition that their power spectrum can grow at scales smaller than those probed by CMB experiments \cite{Guzzetti:2016mkm}. 

Another proposal for measuring at high precision the chirality of PGWs can be found in \cite{Masui:2017fzw}, where it has been claimed that, with futuristic measurements of the 21 cm surveys, one can constrain the chirality at the few percentage level. However, the effective ability of galaxy surveys in building precision maps of the 3D galaxy shape makes this proposal still at a very preliminary stage. Also in \cite{Biagetti:2020lpx} possible detection prospects of parity breaking signatures in the primordial Universe from the 2D galaxy shear power spectrum have been considered: similarly to the case of CMB observations, detection of parity breaking in the near future with this method seems challenging due to instrumental and observational noise.

Thus, given the current experiments, measuring the effects of parity breaking modifications of Einstein  gravity just relying on the power spectrum statistics seems very challenging. As we already mentioned above with a specific example, another important aspect to keep in mind is the high degeneracy regarding the effects of modified gravity operators on the tensor power spectrum. In fact, the final predictions for the level of chirality turn out to be proportional to a ratio $H/M_{\text{PV}}$, with $H$ denoting the Hubble parameter during inflation and $M_{\text{PV}}$ the characteristic energy scale of the parity-violating theories, which is however unknown in the absence of a complete theory of quantum gravity.
This makes all these models indistinguishable from the observational point of view just taking into consideration the power spectrum statistics.    

Therefore, it is interesting and crucial to investigate parity breaking signatures that arise in higher-order correlators, such as the primordial bispectra. In fact, it is well known that the latter contain features (like the shape function) that may be characteristic of a given inflationary model (see, e.g., \cite{Bartolo:2004if,Babich:2004gb,Fergusson:2008ra,Akrami:2019izv}), removing any kind of degeneracy. For instance, in \cite{Maldacena:2011nz,Soda:2011am,Shiraishi:2011st,Huang:2013epa,Zhu:2013fja,Bordin:2017hal,Bartolo:2017szm,Cordova:2017zej,Bartolo:2018elp,Bordin:2020eui} possible parity violating signatures in tensor non-Gaussianity due to modified gravity operators have already been considered.

In this paper we will extend these studies making an original analysis of the effects of new parity violating operators explored in \cite{Crisostomi:2017ugk} on the graviton bispectrum. 
In particular we show that, in the de Sitter limit and assuming constant couplings, the graviton self-interactions contribute to the non-linear graviton wave-function only via a pure phase, thus not affecting the graviton bispectrum. On the other hand we show that these tensor bispectra get interesting features only when we take into account the time dependence of the coupling functions, giving rise to equilateral and squeezed-type bispectra. We discuss the implication of these operators on the so-called ``consistency relation'' for tensor bispectra and the observational prospects for detecting these parity-breaking signatures in the primordial bispectra through experiments focused on the detection of the CMB polarization (like, e.g., the LiteBIRD experiment~\cite{Matsumura:2013aja,Matsumura:2020}).

The paper is organized as follows: in Sec. \ref{CST theories} we will introduce the parity breaking operators that we will study for the rest of the paper. In Sec. \ref{power_spectrum_sec} we will make a review of the effects of these operators on the tensor power spectrum statistics. In Sec. \ref{bispectrum_tensor_sec}, which contains our original computations, we will make an analysis of the parity breaking signatures of these operators in the graviton bispectrum. We will also discuss their original features and observational prospects in light of the recent forecasts on tensor non-Gaussianity from CMB polarization data. Finally, in Sec. \ref{conclusions}, we will draw our conclusions.

\section{Chiral scalar-tensor theories with higher-order derivatives}
\label{CST theories}

In this section we introduce chiral scalar-tensor theories with higher-order derivatives as proposed in \cite{Crisostomi:2017ugk}. These consist in parity breaking covariant terms having more derivatives with respect to both the Einstein-Hilbert term and the so-called gravitational Chern-Simons term, the latter being the full covariant parity breaking operator with less derivatives (see, e.g., \cite{Weinberg:2008hq,Jackiw:2003pm}). The action of these parity-breaking theories has the following form
\begin{align}\label{eq:action_Langlois}
S =  \int d^4 x \sqrt{-g}\left[\frac{M_{Pl}^2}{2}R+\mathcal{L}_{\rm PV}+\mathcal{L}_{\phi}\right]  \, ,
\end{align}
where, as usual, $g = det\left[g_{\mu \nu}\right]$, $M_{Pl}^2 = (8 \pi G)^{-1}$ is the reduced Planck mass, $R$ is the Ricci scalar, $\mathcal{L}_{\rm PV}$ is a Lagrangian containing parity-violating operators, and $\mathcal{L}_\phi$ is the Lagrangian for a scalar degree of freedom, which is assumed to be non-minimally coupled to gravity. Since we are interested in the effects of these parity breaking theories during inflation, we will assume $\phi$ to play the role of the inflaton field, with $\mathcal{L}_\phi$  being the following Lagrangian
\begin{align}
\mathcal{L}_\phi =  -\frac{1}{2} g^{\mu \nu} \partial_\mu \phi \partial_\nu \phi -V(\phi) \, ,
\end{align}
where $V(\phi)$ denotes the (slow-roll) potential of the inflaton field. The parity-violating Lagrangian of the theory can be written as the sum of two pieces 
\begin{align}
\mathcal{L}_{\rm PV} = \mathcal{L}_{\rm PV1} + \mathcal{L}_{\rm PV2} \, ,
\end{align}
where $\mathcal{L}_{\rm PV1}$ contains up to only first derivatives of the scalar field and is given by \cite{Crisostomi:2017ugk}
\begin{align} \label{eq:LPV1}
\mathcal{L}_{\rm PV1} =& \sum_{A=1}^4  a_{A} L_{A}\, ,
\end{align}
where\footnote{Here we are implicitly assuming that the higher derivative operators are suppressed by corresponding powers of the Planck mass, $M_{Pl}$. We will reintroduce these factors explicitly in the next sections.
}
\begin{align}
L_1 &= \varepsilon^{\mu\nu\alpha \beta} R_{\alpha \beta \rho \sigma} R_{\mu \nu\; \lambda}^{\; \; \;\rho} \phi^\sigma \phi^\lambda \, , \qquad \qquad\qquad \qquad L_3 = \varepsilon^{\mu\nu\alpha \beta} R_{\alpha \beta \rho \sigma} R^{\sigma}_{\;\; \nu} \phi^\rho \phi_\mu \, ,\nonumber\\
L_2 &=  \varepsilon^{\mu\nu\alpha \beta} R_{\alpha \beta \rho \sigma} R_{\mu \lambda }^{\; \; \;\rho \sigma} \phi_\nu \phi^\lambda \, , \qquad \qquad\qquad \qquad  L_4 =  \varepsilon^{\mu\nu\rho\sigma} R_{\rho\sigma \alpha\beta} R^{\alpha \beta}_{\;\;\;\; \mu\nu} \phi^\lambda \phi_\lambda \, ,
\label{opLPV1}
\end{align}
where $\varepsilon_{\rho \sigma \alpha \beta}$ is the covariant Levi-Civita tensor defined in terms of the antisymmetric Levi-Civita symbol $\epsilon^{\rho \sigma \alpha \beta}$ as $\varepsilon^{\rho \sigma \alpha \beta}=\epsilon^{\rho \sigma \alpha \beta}/\sqrt{-g}$, and $\phi^\mu = \nabla^\mu \phi$ with  $\nabla^\mu$ denoting the covariant derivative.
Notice that the couplings $a_A$ in \eqref{eq:LPV1} are generic functions of the scalar field and its kinetic term, i.e. $a_A = a_{A}(\phi, \phi^\mu \phi_\mu)$.

In \cite{Crisostomi:2017ugk} it has been shown that in the so-called unitary gauge, where the scalar field is homogeneous, these operators do not introduce the Ostrogradsky (unstable) modes under the constraint
\begin{equation} \label{eq:constraint_a}
4a_1+2 a_2+a_3 +8 a_4=0 \, ,
\end{equation}
that leaves only 3 independent coefficients. 

On the other hand, the term $\mathcal{L}_{\rm PV2}$, which includes also second-order derivatives of the scalar field, reads \cite{Crisostomi:2017ugk}
\begin{align} \label{eq:LPV2}
\mathcal{L}_{\rm PV2} &= \sum_{A=1}^7 b_{A} M_{A}\, ,
\end{align}
where
\begin{align}
M_1 &= \varepsilon^{\mu\nu \alpha \beta} R_{\alpha \beta \rho\sigma} \phi^\rho \phi_\mu \phi^\sigma_\nu \, , \qquad \qquad\qquad \qquad  M_4 = \varepsilon^{\mu\nu \alpha \beta} R_{\alpha \beta \rho\sigma} \phi_\nu \phi_\mu^\rho \phi^\sigma_\lambda \phi^\lambda, \nonumber\\
M_2 &= \varepsilon^{\mu\nu \alpha \beta} R_{\alpha \beta \rho\sigma} \phi^\rho_\mu \phi^\sigma_\nu \, , \qquad \qquad\qquad \qquad\,\,\,\,\,\, M_5 = \varepsilon^{\mu\nu \alpha \beta} R_{\alpha \rho\sigma \lambda } \phi^\rho \phi_\beta \phi^\sigma_\mu \phi^\lambda_\nu,  \nonumber\\
M_3 &= \varepsilon^{\mu\nu \alpha \beta} R_{\alpha \beta \rho\sigma} \phi^\sigma \phi^\rho_\mu \phi^\lambda_\nu \phi_\lambda\, , \qquad \qquad\qquad\,\,\,\,\,   M_6 = \varepsilon^{\mu\nu \alpha \beta} R_{\beta \gamma} \phi_\alpha \phi^\gamma_\mu \phi^\lambda_\nu \phi^\lambda, \nonumber\\
M_7 &= (\Box \phi) M_1 \, ,
\label{opLPV2}
\end{align}
where in this case $\phi^{\sigma}_\nu = \nabla^\sigma \nabla_\nu \phi$ and $b_A = b_{A}(\phi, \phi^\mu \phi_\mu)$. 

Here, in order to avoid the Ostrogradsky modes in the unitary gauge, the following conditions have to be imposed \cite{Crisostomi:2017ugk}
\begin{equation} \label{eq:constraint_b}
 b_7 = 0 \, , \qquad  b_6 = 2(b_4 + b_5) \,,  \qquad b_2 =-A_*^2(b_3 -b_4)/2 \, ,
\end{equation}
where $A_* = \dot \phi(t)/N$ and $N$ is the so-called lapse function of the spacetime. In this case, we are left with 4 independent coefficients.

In the following sections we will analyze signatures of these operators on tensor perturbations during slow-roll inflation, reviewing results about power spectra statistics and making an original analysis about tensor non-Gaussianity.

In order to do the computations, we will adopt the so-called Arnowitt-Deser-Misner (ADM) formalism for the perturbed Friedmann-Lemaitre-Robertson-Walker (FLRW) metric (see e.g. \cite{Arnowitt:1962hi,Bardeen:1980kt,Maldacena:2002vr}), where the metric reads
\begin{equation}
ds^2 = -(N^2 - N_i N^i)\, dt^2 + N_i \,dx^i dt + h_{ij} \,dx^i dx^j \, ,
\end{equation}
where $N$ and $N_i$ are the so-called lapse and shift functions respectively and $h_{ij}$ is the three-dimensional spatial metric. Focusing only on transverse and traceless tensor perturbations $\gamma_{ij}$, the (non-linear) perturbed spatial metric reads \cite{Salopek:1990jq,Maldacena:2002vr}
\begin{align}
&h_{ij} = a^2\left[\exp \gamma \right]_{ij} \, , \qquad  {\gamma_i}^i = 0 \, , \quad \partial_i \gamma^{ij} =0 \, ,  \nonumber \\
&\left[\exp \gamma \right]_{ij}  = \delta_{ij} + \gamma_{ij} + \frac{1}{2!} \gamma_{ik} {\gamma^{k}}_{j} + ... \, ,
\end{align}
where $a(t)$ denotes the scale factor of the Universe.
Moreover, in standard gravity $N$ and $N_i$ are non-dynamical fields that can be removed in the final action after solving their algebraic Euler-Lagrange equations in terms of dynamical fields. In general, in a modified gravity setting, the equations of motion for $N$ and $N_i$ are modified, with the possibility to have an increased number of dynamical degrees of freedom. However, it has been shown in \cite{Crisostomi:2017ugk} that in the unitary gauge and under the constraints \eqref{eq:constraint_a} and \eqref{eq:constraint_b} no new degrees of freedom are introduced in the theory with respect to standard gravity.      

In App. \ref{App:Interaction-Hamiltonian} we report the expressions of the Lagrangians \eqref{eq:LPV1}-\eqref{eq:LPV2} in the unitary gauge within the ADM formalism, as derived in \cite{Crisostomi:2017ugk}. We will assume these to be the fundamental Lagrangians defining the theories that we will study. Indeed, as commented in \cite{Crisostomi:2017ugk}, because of the presence of the Ostrogradsky modes in the original theories \eqref{eq:LPV1}-\eqref{eq:LPV2}, there are two possible approaches to follow: the first one is to consider these theories as low energy effective field theories valid up to the energy scale at which the Ostrogradsky modes appear; the second possibility is to restrict the theories to the unitary gauge, with the additional constraints \eqref{eq:constraint_a}-\eqref{eq:constraint_b}, and treat them as new Lorentz breaking (and parity violating) theories. The latter is the approach adopted in Ref. \cite{Crisostomi:2017ugk} and which we will also follow in the rest of the paper.  Notice that the two Lagrangians \eqref{eq:PV1-action-ADM}-\eqref{eq:PV2-action-ADM} do not contain any higher order time derivative of the metric, but only higher order space derivatives. Because of this fact, these theories break Lorentz invariance similarly to what happens in Horava-Lifshitz gravity \cite{Horava:2009uw,Wang:2017brl}. As we will see in the rest of the paper, this feature will have important consequences on the phenomenology of the models under scrutiny, both in the propagation of PGWs (leading to a speed of propagation of tensor modes different from the speed of light during inflation) and in the predictions for primordial bispectra. 

Before analyzing in details the effects of these parity-breaking operators on primordial tensor modes, some comments are in order regarding inflation within these modified gravity theories. The first thing to consider when introducing some modifications to gravity is how the dynamics at the background level is affected. In our specific case, it is easy to check that the new operators have no effects on the background dynamics of inflation, that is the same as in single field slow-roll models within general relativity. At the perturbation level, instead, rotational invariance implies that $N$-point correlation functions of scalar perturbations can have parity-odd signals only for $N \ge 4$, i.e. the trispectrum ($N=4$) is the lowest order correlator involving only scalars that can manifest parity-breaking signatures \cite{Shiraishi:2016mok}. Thus, the power spectrum of scalar perturbations has the usual expression that is obtained within general relativity.

Finally let us recall a more technical point, which is however important for the following computations: if one is not interested in expanding the actions beyond cubic order in the perturbations, it is sufficient to find the expressions of the lapse function $N$ and the shift vector $N_i$ at first order only. However, since we are focusing only on tensor fluctuation modes and it is not possible to have first order perturbations in $N$ and $N_i$ including only tensor perturbations, we are left solely with their zero-order value, namely 
\begin{align}
N=1 \, ,\qquad N_i=0 \, .
\end{align}

\section{Chirality in primordial tensor power-spectra} \label{power_spectrum_sec}

In this section we are going to review the effects of higher order operators introduced in \cite{Crisostomi:2017ugk} on the dynamics of PGWs, with a particular focus on the parity breaking signatures that arise in the primordial tensor power spectrum. This has been studied in \cite{Qiao:2019hkz} (see also \cite{Qiao:2019wsh,Nishizawa:2018srh,Gao:2019liu} for an analysis of the propagation of gravitational waves in the late-time Universe within these parity-breaking theories).
We first start by recalling the Fourier expansion of PGWs, that is given by
\begin{equation}
\gamma_{ij}(\mathbf{x},t)=\int\frac{d^3k}{(2\pi)^3}\sum_{s=L,R}\gamma_s(\mathbf{k},t)\epsilon^{(s)}_{ij}(\mathbf{k})e^{i\mathbf{k}\cdot\mathbf{x}} \, ,
\end{equation}
where $\gamma_s(\mathbf{k},t)$ are the mode function of primordial tensor modes, $s$ being the polarization index, and $\epsilon^{(s)}_{ij}(\mathbf{k})$ are the polarization tensor in the chiral basis, i.e. in the basis of L and R circular polarization states. This is defined in terms of the more common $+$ and $\times$ basis as
\begin{equation}
\epsilon^R_{ij}(\mathbf{k})=\frac{\epsilon^{+}_{ij}(\mathbf{k})+ i\epsilon^{\times}_{ij}(\mathbf{k})}{\sqrt{2}} \, ,\quad \epsilon^L_{ij}(\mathbf{k})=\frac{\epsilon^{+}_{ij}(\mathbf{k})- i\epsilon^{\times}_{ij}(\mathbf{k})}{\sqrt{2}} \, ,
\end{equation}
and the mode functions in the two different bases are related by
\begin{equation}
\gamma_R(\mathbf{k},t)=\frac{\gamma_+(\mathbf{k},t)- i\gamma_{\times}(\mathbf{k},t)}{\sqrt{2}} \, ,\quad \gamma_L(\mathbf{k},t)=\frac{\gamma_+(\mathbf{k},t)+ i\gamma_{\times}(\mathbf{k},t)}{\sqrt{2}} \, .
\end{equation}
This decomposition into circular polarization states is particularly useful when studying parity violating theories. Indeed, while the $+$ and $\times$ polarization states are mixed by the parity violating terms, the equations of motion for the L and R polarization modes are decoupled. One can prove that the following relations hold (see, e.g., \cite{Alexander:2004wk})
\begin{align}
\epsilon_{ij}^{L}(\mathbf{k})\epsilon_{L}^{ij}(\mathbf{k})&=\epsilon_{ij}^{R}(\mathbf{k})\epsilon_{R}^{ij}(\mathbf{k}) = 0 \, , \nonumber\\
\epsilon_{ij}^L(\mathbf{k})\epsilon_R^{ij}(\mathbf{k})&=2 \, ,\nonumber\\
\epsilon_{ij}^{R}(-\mathbf{k})&=\epsilon_{ij}^{L}(\mathbf{k}) \, ,\nonumber\\
\epsilon^{(s)*}_{ij}(-\mathbf{k})&=\epsilon^{(s)}_{ij}(\mathbf{k}),\nonumber\\
\gamma_{L}(-\mathbf{k})&=\gamma_{R}^*(\mathbf{k}) \, ,\nonumber\\
k_l\epsilon^{mlj}{\epsilon_j^{(s) j}}(\mathbf{k})&=-i\lambda_sk\epsilon^{(s) im}(\mathbf{k}) \, ,\label{eq:circ_identities}
\end{align}
where $\lambda_R=+1$ and $\lambda_L=-1$, and $\epsilon^{mlj}$ with 3 Latin indices denotes the Levi-Civita anti-symmetric symbol. We will make an extensive use of these relations throughout all of the paper.

The first step required to compute the power spectrum of PGWs is to expand the Lagrangians introduced in the previous section at second order in tensor perturbations. 
Working at leading order in slow-roll parameters, the action derived by Lagrangian PV1 \eqref{eq:PV1-action-ADM} (including the contribution from standard gravity) at quadratic order in tensor perturbations is given by
\begin{equation}
\label{PV1-quadratic}
S_{\gamma\gamma}^{\text{PV1}}=\sum_{s=L,R}\int d\tau\int\frac{d^3k}{(2\pi)^3}\Big[A_{T,s}^2|\gamma'_s(\mathbf{k},\tau)|^2-B_{T,s}^2k^2|\gamma_s(\mathbf{k},\tau)|^2\Big] \, ,
\end{equation}
where the prime denotes a derivative with respect to the conformal time $\tau$ and we have defined 
\begin{equation}
\label{PV1-A-B}
A_{T,s}^2\equiv\frac{M_{Pl}^2}{2}a^2\left(1-\lambda_s\frac{k_{phys}}{M_{\text{PV1}}}\right),\quad B_{T,s}^2\equiv\frac{M_{Pl}^2}{2}a^2\left[1-\frac{4}{M_{Pl}^6}\frac{\dot{\phi}^2}{a}(\dot{f}+\dot{g})\lambda_sk\right] \, ,
\end{equation}
with
\begin{equation}
\label{MPV1}
M_{\text{PV1}}\equiv\frac{M_{Pl}^6}{8}\frac{1}{\dot{\phi}^2}\frac{1}{(f+g)H} \, ,
\end{equation}
and
\begin{equation}
\label{PV1-f-g}
f\equiv a_1+\frac{a_2}{2}+2a_4 \, ,\qquad g\equiv\frac{a_2}{2}+2a_4 \, ,	
\end{equation}	   
where the dot denotes a derivative with respect to cosmic time.
From Eq. \eqref{PV1-A-B} we realize that the right-handed graviton modes ($\lambda_R=+1$) with a physical wavenumber, $k_{phys}=k/a$, larger than $M_{\text{PV1}}$ get a negative kinetic term, thus becoming unstable. At the quantum level this instability may result in severe problems, since it leads either to a violation of unitarity or to the propagation of negative energy modes forward in time. Since unitarity has to be preserved in order for the theory to make sense, we must admit the presence of particles with negative energies, which means that the energy spectrum is unbounded from below. However, in such a case, the vacuum state would be highly unstable under the decay into particles of positive and negative energies \cite{Dyda:2012rj}.
So, to avoid to deal with this kind of problem we introduce a UV cut-off $\Lambda \le M_{\text{PV1}}$ in the theory and consider only gravitons with $k_{phys}<\Lambda$ at the inset of inflation. Then, since deep inside the horizon the condition $k_{phys}\gg H$ holds, it follows that we must assume $M_{\text{PV1}}\gg H$ during inflation in order for the theory to make sense\footnote{Notice that assuming just $M_{\text{PV1}}  \gtrsim H$ is not enough, as in this case a given physical mode $k_{phys}$ encounters issues near the horizon crossing. Thus, this model can be considered well-defined only when $M_{\text{PV1}}\gg H$, i.e. in the ``effective field theory'' limit, with $M_{\text{PV1}}$ playing the role of scale of new physics.}.

As explained in Sec. \ref{CST theories}, the couplings $a_i$ that enter in the definitions \eqref{PV1-f-g} of $f$ and $g$ are functions of the scalar field and its kinetic term ($=\dot{\phi}^2/2$ in the unitary gauge). This allows us to reabsorb the $\dot{\phi}^2$ terms in Eqs. \eqref{PV1-A-B}-\eqref{MPV1} by defining two new couplings\footnote{Notice that, despite the fact that operators in \eqref{eq:LPV1} contain 6 derivatives, only some of them act on the perturbations, while the others act on the background. Thus, since these theories make sense only in unitary gauge, we are allowed to reabsorb some of the (time) derivatives through a coupling redefinition.}
\begin{equation}
\label{f1g1}
f_1\equiv\frac{\dot{\phi}^2}{M_{Pl}^4}f \, ,\quad g_1\equiv\frac{\dot{\phi}^2}{M_{Pl}^4}g \, ,
\end{equation}
that are still dimensionless like $f$ and $g$.
If we now define the graviton speed as
\begin{equation}
c_{T,s}^2\equiv\frac{B_{T,s}^2}{A_{T,s}^2} \, ,
\end{equation}
the action \eqref{PV1-quadratic} can be rewritten as
\begin{equation}
S_{\gamma\gamma}^{\text{PV1}}=\sum_{s=L,R}\int d\tau\int\frac{d^3k}{(2\pi)^3}A^2_{T,s}\Big[|\gamma'_s(\mathbf{k},\tau)|^2-c_{T,s}^2k^2|\gamma_s(\mathbf{k},\tau)|^2\Big] \, .
\end{equation}
By making the field redefinition
\begin{equation}
\mu_s\equiv A_{T,s}\gamma_s \, ,
\end{equation}	
we can then rewrite the action for the new field as
\begin{equation}
S_{\gamma\gamma}^{\text{PV1}}=\sum_{s=L,R}\int d\tau\int\frac{d^3k}{(2\pi)^3}\left[|\mu'_s(\mathbf{k},\tau)|^2-c^2_{T,s}k^2|\mu_s(\mathbf{k},\tau)|^2+\frac{A''_{T,s}}{A_{T,s}}|\mu_s(\mathbf{k},\tau)|^2\right] \, .
\end{equation}	
Varying this action yields the equations of motion for the fields $\mu_s$, which read
\begin{equation}
\mu''_s+\left(c^2_{T,s}k^2-\frac{A''_{T,s}}{A_{T,s}}\right)\mu_s=0 \, ,
\end{equation}
where the time-dependent effective mass is
\begin{equation}
\label{eq:PV1-effective-mass}
\frac{A_{T,s}''}{A_{T,s}}=\frac{d}{d\tau}\left(\frac{A_{T,s}'}{A_{T,s}}\right)+\left(\frac{A_{T,s}'}{A_{T,s}}\right)^2=\frac{2+3\epsilon}{\tau^2}-\frac{\lambda_sk}{\tau}\frac{H}{M_{\text{PV1}}}+\mathcal{O}\left(\epsilon^2,\frac{H^2}{M_{\text{PV1}}^2},\epsilon\frac{H}{M_{\text{PV1}}}\right) \, ,
\end{equation}	
and $\epsilon$ is a slow-roll parameter, defined as
\begin{equation}
\label{eq:slow-roll-parameter}
\epsilon=\frac{M_{Pl}^2}{2}\left(\frac{V'}{V}\right)^2\simeq\frac{1}{2}\frac{\dot{\phi}^2}{H^2M_{Pl}^2} \, .
\end{equation}
The first term in Eq. \eqref{eq:PV1-effective-mass} is the usual contribution present in slow-roll models of inflation with standard gravity. The second term, instead, is a new contribution that arises similarly in inflationary models with the gravitational Chern-Simons coupling (see \cite{Alexander:2004wk,Satoh:2010ep,Bartolo:2017szm,Mylova:2019jrj}), with the Chern-Simons mass replaced in our case by $M_{\text{PV1}}$.
Thus, the equations of motion for the fields $\mu_s$ are
\begin{equation}
\label{PV1-eom}
\mu''_s+\left(c_{T,s}^2k^2-\frac{\nu_T^2-\frac{1}{4}}{\tau^2}+\lambda_s\frac{k}{\tau}\frac{H}{M_{\text{PV1}}}\right)\mu_s=0 \, ,
\end{equation}	
with
\begin{equation}
\nu_T\simeq\frac{3}{2}+\epsilon \, ,
\end{equation}
and
\begin{equation}
\label{PV1-cT}
c_{T,s}^2\simeq 1 - \lambda_s k\frac{H}{M_{\text{PV1}}}\tau 
\end{equation}
at leading order in slow-roll dynamics and in the ratio $H/M_{\text{PV1}}$.

From Eq. \eqref{PV1-eom} we realize that, due to the presence of the higher derivative operators, the speed of propagation of tensor modes is modified in this model, since $c_{T,s}\ne 1$; this is actually already evident from the action \eqref{PV1-quadratic}, since the time and space derivatives of the field are multiplied by different functions.
In particular, from \eqref{PV1-cT} we notice that the two circular polarization states propagate with a different speed during inflation, being this dependent on the polarization index $s$: this is the so-called velocity birefringence phenomenon and it is the main difference at quadratic level with respect to the case with Chern-Simons gravity. We also stress that the speed of tensor modes is not constant, but varies with time during inflation. This is a peculiar feature of this kind of models.
Notice that in the limit where $f_1=g_1=0$ ($M_{\text{PV1}} = \infty$) we recover the usual result, $c_{T,s}=1$, as expected.

There is a further important feature that arises from Eq. \eqref{PV1-cT}: during inflation one of the two polarization states is superluminal (i.e. $c_{T,s}>1$), while the other one is subluminal. This was already noticed e.g. in \cite{Nishizawa:2018srh} and is a phenomenological manifestation of the breaking of Lorentz invariance that occurs in this model. Notice that such an invariance is recovered at the end of inflation.

We can now canonically quantize the fields $\mu_s$ by expanding them in terms of the creation and annihilation operators 
\begin{equation}
\hat{\mu}_s(\mathbf{k},\tau)=u_s(k,\tau)\hat{a}_s(\mathbf{k})+u^*_s(k,\tau)\hat{a}^{\dagger}_s(-\mathbf{k}) \, .
\end{equation}
The creation and annihilation operators satisfy the equal time commutation relations
\begin{equation}
[\hat{a}_s(\mathbf{k}),\hat{a}^{\dagger}_{s'}(\mathbf{k}')]=(2\pi)^3\delta^{(3)}(\mathbf{k}-\mathbf{k}')\delta_{ss'} \, ,\quad[\hat{a}_s(\mathbf{k}),\hat{a}_{s'}(\mathbf{k}')]=0=[\hat{a}_s^{\dagger}(\mathbf{k}),\hat{a}_{s'}^{\dagger}(\mathbf{k}')] \, ,
\end{equation}
and act on the vacuum state as
\begin{equation}
\hat{a}_s|0\rangle=0 \, ,\quad\langle0|\hat{a}_s^{\dagger}=0 \, .
\end{equation}
The equations of motion for the mode functions $u_s$ follow straightforwardly from Eq. \eqref{PV1-eom} and read
\begin{equation}
\label{PV1-eom-mode-functions}
u_s''+\left[k^2 \left(1-\lambda_s k\frac{H}{M_{\text{PV1}}}\tau\right) - \frac{\nu_T^2-\frac{1}{4}}{\tau^2}+\lambda_s\frac{k}{\tau}\frac{H}{M_{\text{PV1}}}\right] u_s=0 \, .
\end{equation}
Now, this equation has the same form as Eq.~(4.11) of \cite{Qiao:2019hkz} when taking $c_2 = 0$ and $c_1 \epsilon_* = H/M_{\text{PV1}}$. As shown in \cite{Qiao:2019hkz}, equations of the kind of \eqref{PV1-eom-mode-functions} admit an approximate analytical solution in terms of Airy functions \cite{NIST:DLMF}
\begin{align}\label{eq_usolution}
u_s(y) = \alpha \left(\frac{\xi(y)}{g(y)}\right)^{1/4} \textrm{Ai}(\xi) + \beta \left(\frac{\xi(y)}{g(y)}\right)^{1/4} \textrm{Bi}(\xi) \, ,
\end{align}
where $\alpha$ and $\beta$ are two integration constants, $y = - k \tau$ and the functions $\xi(y)$ and $g(y)$ in our conventions are given by
\begin{align}
g(y) = \frac{\nu_T^2}{y^2} - 1 - \lambda_s y \frac{H}{M_{\text{PV1}}} + \lambda_s \frac{H}{M_{\text{PV1}}} \frac{1}{y}  \, ,    
\end{align}
and
\begin{align}
\xi(y) = \begin{cases}
\left(-\frac{3}{2} \int_{y_0^s}^y \sqrt{g(y')} \, dy' \right)^{2/3} \qquad\thinspace y \leq y_0^s \, ,\\
- \left(\frac{3}{2} \int_{y_0^s}^y \sqrt{g(y')} \, dy' \right)^{2/3} \qquad y \geq y_0^s \, ,
\end{cases}
\end{align}
with 
\begin{align}
y_0^s= -\frac{1-2^{1/3}\left[1+3\left(\frac{H}{M_{\text{PV1}}}\right)^2 \right]/Y -2^{-1/3}Y}{3\lambda_s \frac{H}{M_{\text{PV1}}}} \, ,
\end{align}
where
\begin{align}
Y=&\left(Y_1+\sqrt{-4\left[1+3\left(\frac{H}{M_{\text{PV1}}}\right)^2\right]^3+Y_1^2}\right)^{1/3} \, ,\\
Y_1=&-2+27 \nu_T^2 \left(\frac{H}{M_{\text{PV1}}}\right)^2-9 \left(\frac{H}{M_{\text{PV1}}}\right)^2 \, .
\end{align}
The final solution for $u_s(y)$ is found by matching the sub-horizon limit ($y \rightarrow \infty$) of \eqref{eq_usolution} with the following initial condition
\begin{align} \label{eq:adiab_init}
\lim_{y \to +\infty} u_s(y)= \sqrt{\frac{1}{2 \omega_k}}\exp\left(-i \int^\tau_{\tau_i}  \omega_k \, d\tau'\right) \,,
\end{align}
that physically corresponds to the assumption that the Universe was initially in an adiabatic vacuum state. Here $\omega_k = k \sqrt{- g(-k \tau)}$ denotes the dispersion relation of PGWs that can be read off by Eq. \eqref{PV1-eom-mode-functions}. Notice that \eqref{eq:adiab_init} corresponds to the usual Bunch-Davies initial vacuum state in the limit in which $\omega_k = k$. By doing this matching we find \cite{Qiao:2019hkz} 
\begin{align}
\alpha = \sqrt{\frac{\pi}{2 k}} e^{i \pi/4} \, , \qquad \qquad  \beta = i \sqrt{\frac{\pi}{2 k}} e^{i \pi/4} \, ,
\end{align}
that fixes our solution. In \cite{Qiao:2019hkz} it has been shown that this analytical approach is in optimal agreement with exact numerical solutions. 

We can now derive the super-horizon power spectra for the two circular polarization modes of tensor perturbations, which are defined as
\begin{equation}
P_T^L=2\frac{|u_L(y)_{y \ll 1}|^2}{A_{T,L}^2} \, ,\qquad P_T^R=2\frac{|u_R(y)_{y \ll 1}|^2}{A_{T,R}^2} \, .
\end{equation}
At leading order in slow-roll the final result reads \cite{Qiao:2019hkz}
\begin{equation}
\label{PV1-PS}
P_T^L= \frac{P_T}{2} \exp\left[\frac{\pi}{16}\frac{H}{M_{\text{PV1}}}\right] \, ,\qquad P_T^R= \frac{P_T}{2} \exp\left[-\frac{\pi}{16}\frac{H}{M_{\text{PV1}}}\right] \, ,
\end{equation}
where here $P_T$ denotes the total tensor power spectrum as predicted in general relativity. 

Now, the level of parity violation in the power spectra of primordial tensor modes can be quantified by means of the chirality parameter $\chi$, which is defined as the relative difference between the power spectra of right and left polarization modes. The leading-order contributions to the chirality can be computed by Taylor-expanding the exponentials in Eqs. \eqref{PV1-PS}, since $H/M_{\text{PV1}}\ll1$. Thus, we find
\begin{equation}
\label{PV1-chi}
\chi\equiv\frac{P_T^R-P_T^L}{P_T^R+P_T^L}= - \frac{\pi}{16}\frac{H}{M_{\text{PV1}}} \, .
\end{equation} 
From Eq. \eqref{PV1-chi} it is clear that, just like it happens with the gravitational Chern-Simons coupling \cite{Alexander:2004wk,Satoh:2010ep,Bartolo:2017szm,Mylova:2019jrj}, the chirality is suppressed by the requirement of dealing with an effective field theory, i.e. $H/M_{PV1}\ll 1$.
This indeed does not come as a surprise, since, as already mentioned, in this case the equations of motion~\eqref{PV1-eom} are similar to the case of a Chern-Simons coupling. 

Another interesting quantity to compute is the modifications to the tensor-to-scalar ratio with respect to the result obtained within standard gravity. The total dimensionless tensor power spectrum can be written as
\begin{equation}
\Delta_T^{\text{PV1}}=\Delta_T^R+\Delta_T^L=\Delta_T\left[1+\frac{\pi^2}{256}\left(\frac{H}{M_{\text{PV1}}}\right)^2\right]=\Delta_T\left(1+\chi^2\right) \, .   
\end{equation}
Since, as discussed in Sec. \ref{CST theories}, the scalar power spectrum does not receive any contribution from the parity-violating operators, the tensor-to-scalar ratio can be readily computed as
\begin{equation}
\label{eq:PV1-tensor-to-scalar}
r_{\text{PV1}}\equiv\frac{\Delta_T^{\text{PV1}}}{\Delta_S}=r\left(1+\chi^2\right) \, ,    
\end{equation}
where $r$ is the tensor-to-scalar ratio obtained in slow-roll models without the parity-breaking operators. We can clearly see that, since the chirality $\chi$ is $\ll 1$ in this model, the correction to $r$ is suppressed. 

The presence of the new operators induces some corrections also to the spectral index of tensor perturbations, that quantifies how the amplitude of the fluctuations varies with the scale. This has the following expression
\begin{equation}
\label{eq:PV1-tensor-tilt}
n_T\equiv\frac{d\ln\Delta_T^{\text{PV1}}}{d\ln k}\simeq-2\epsilon+\frac{\pi^2}{128}\left(\frac{H}{M_{\text{PV1}}}\right)\left[-2\epsilon\left(\frac{H}{M_{\text{PV1}}}\right)-\frac{\dot{M}_{\text{PV1}}}{M_{\text{PV1}}^2}\right] \, .
\end{equation}
Thus, even in this case the corrections to the standard result ($n_T=-2\epsilon$) are in general  small\footnote{
Indeed, it is easy to show that
\begin{equation}
\frac{\dot{M}_{\text{PV1}}}{M_{\text{PV1}}^2}\simeq\epsilon\left(\frac{H}{M_{\text{PV1}}}\right)-\sqrt{2\epsilon}M_{Pl}\left(\frac{H}{M_{\text{PV1}}}\right)\frac{(\partial f_1/\partial\phi)+(\partial g_1/\partial\phi)}{f_1+g_1} \, .
\end{equation}
}, making PGWs predicted within this model far from the reach of GW interferometers.

Now we want to make a similar analysis for the Lagrangian PV2, given by Eq.~(\ref{eq:PV2-action-ADM}). At second order in tensor perturbations it has the following form
\begin{equation}
\label{PV2-quadratic}
S_{\text{PV2}}^{\gamma\gamma}=\sum_{s=L,R}\int d\tau\int\frac{d^3k}{(2\pi)^3}\left[\tilde{A}_{T,s}^2|\gamma'_s(\mathbf{k},\tau)|^2-\frac{M_{Pl}^2}{2}a^2k^2|\gamma_s(\mathbf{k},\tau)|^2\right] \, ,
\end{equation}
where we have defined
\begin{equation}
\label{ATMPv2}
\tilde{A}_{T,s}^2\equiv\frac{M_{Pl}^2}{2}a^2
\left(1-\lambda_s\frac{k_{phys}}{M_{\text{PV2}}}\right) \, ,
\end{equation}
with
\begin{equation}
M_{\text{PV2}}\equiv\frac{M_{Pl}}{2}\left(\tilde{b}_1-b\frac{H}{M_{Pl}}\right)^{-1} \, .
\end{equation}
Just like in the previous case, we have reabsorbed the powers of $\dot{\phi}^2$ by defining two new couplings $\tilde{b}_1$ and $b$ as
\begin{equation}
\label{b1-tilde}
\tilde{b}_1\equiv\frac{\dot{\phi}^3}{M_{Pl}^6}b_1 \, ,\quad b\equiv \frac{\dot{\phi}^4}{M_{Pl}^8}(b_4+b_5-b_3) \, ,
\end{equation}
where $b_1,b_3,b_4$ and $b_5$ are the independent couplings of the model. 
The energy scale $M_{\text{PV2}}$ has been introduced for the same reason as $M_{\text{PV1}}$ in the case with only first derivatives of the scalar field: it represents the energy scale at which the right-handed graviton modes acquire a negative kinetic term, thus becoming unstable. Proceeding as in the previous case, we introduce a UV cut-off $\Lambda \le M_{\text{PV2}}$ imposing that $k_{phys}<\Lambda$. By requiring also that the modes started deep inside the horizon, it follows that $H/M_{\text{PV2}}\ll 1$.

By defining the new field $\mu_s\equiv \tilde{A}_{T,s}\gamma_s$ and repeating the same steps as in the previous case, we obtain the following equations of motion 
\begin{equation}
\label{PV2-eom}
\mu_s''+\left(\tilde{c}_{T,s}^2k^2-\frac{\nu_T^2-\frac{1}{4}}{\tau^2}+\lambda_s\frac{k}{\tau}\frac{H}{M_{\text{PV2}}}\right)\mu_s=0 \, ,
\end{equation}
where also in this case
\begin{equation}
\nu_T\simeq\frac{3}{2}+\epsilon
\end{equation}
holds at leading order in slow-roll parameters. The speed of propagation of tensor modes during inflation can be written at leading order in slow-roll and in the ratio $H/M_{\text{PV2}}$ as
\begin{equation}
\tilde{c}_{T,s}^2\simeq 1 - \lambda_s k\frac{H}{M_{\text{PV2}}}\tau \, .
\end{equation}
If we then canonically quantize the field $\mu_s$ as
\begin{equation}
\hat{\mu}_s(\mathbf{k},\tau)=u_s(k,\tau)\hat{a}_s(\mathbf{k})+u^*_s(k,\tau)\hat{a}^{\dagger}_s(-\mathbf{k}) \, ,
\end{equation}
we can immediately write down the equations of motion for the mode functions $u_s$, which read
\begin{equation}
\label{PV2-eom-mode-functions}
u_s''+\left[k^2 \left(1 - \lambda_s k\frac{H}{M_{\text{PV2}}}\tau \right)-\frac{\nu_T^2-\frac{1}{4}}{\tau^2}+\lambda_s\frac{k}{\tau}\frac{H}{M_{\text{PV2}}}\right] u_s=0 \, .
\end{equation}
This is basically the same equation as Eq. \eqref{PV1-eom-mode-functions}, with $M_{\text{PV2}}$ replacing $M_{\text{PV1}}$. We can thus write the solution of Eq. \eqref{PV2-eom-mode-functions} in terms of the Airy functions and the integration constants are again fixed by imposing the adiabatic initial condition \eqref{eq:adiab_init}. We can then compute the leading order contribution to the chirality parameter $\chi$, which takes the following form
\begin{equation}
\label{PV2-chi}
\chi= - \frac{\pi}{16}\frac{H}{M_{\text{PV2}}}  \, .
\end{equation} 
As in the PV1 model, the chirality of PGWs is suppressed since $H/M_{\text{PV2}}\ll 1$. Analogous considerations hold for the tensor-to-scalar ratio and the spectral index of tensor perturbations, that have the same expressions as in Eqs. \eqref{eq:PV1-tensor-to-scalar}-\eqref{eq:PV1-tensor-tilt} with $M_{\text{PV2}}$ replacing $M_{\text{PV1}}$.

At this point, we want to make a brief comment about the observability of this signature: as emphasized in the introduction, the CMB $EB$ and $TB$ angular cross-correlators are able to probe parity breaking in the primordial Universe only for models predicting maximum chirality, i.e. $\chi \simeq 1$ \cite{Gerbino:2016mqb}, while the models under considerations predict $\chi \ll 1$. Moreover, as we have already discussed, in general we do not expect a significant modification of the standard slow-roll models tensor tilt (see Eq. \eqref{eq:PV1-tensor-tilt}), making in general difficult to probe these models with forthcoming interferometers.

Thus, measuring the linear effects of these parity breaking operators seems very challenging. Another important aspect to keep in mind is the high degeneracy regarding their signatures on the primordial power spectra. In fact, the final predictions for the level of chirality \eqref{PV1-chi} and \eqref{PV2-chi} are equivalent apart for a redefinition of the $M_{\text{PV}}$ scale, which is unknown in the absence of a more fundamental theory able to predict its value. This makes these models indistinguishable just taking into consideration the power spectrum statistics.    

Therefore, it is interesting and crucial to investigate the kind of parity breaking signatures that arise in higher-order correlators, like, e.g., the graviton bispectrum. In fact, the latter contains features that may be characteristic of a given inflationary model, removing any kind of degeneracy. For this reason, in the next section we are going to make an original study of the chirality in the primordial tensor bispectra provided by the set of operators in \eqref{eq:LPV1} and \eqref{eq:LPV2}.

\section{Chirality in primordial tensor bispectra} \label{bispectrum_tensor_sec}

Before entering into the details of the computations, let us first recall the basic definition of the bispectrum, which is the Fourier transform of the three-point correlation function. Given three perturbation fields $\delta_1(\mathbf{x},t)$, $\delta_2(\mathbf{x},t)$ and $\delta_3(\mathbf{x},t)$, the bispectrum $B(k_1,k_2,k_3)$ is defined through the relation
\begin{equation}
\langle\delta_1(\mathbf{k_1})\delta_2(\mathbf{k_2})\delta_3(\mathbf{k_3})\rangle=(2\pi)^3\delta^{(3)}(\mathbf{k_1}+\mathbf{k_2}+\mathbf{k_3})B(k_1,k_2,k_3)\, .
\end{equation}
The Dirac delta enforces the invariance under spatial translations and implies that the wave vectors $\mathbf{k_1}$, $\mathbf{k_2}$ and $\mathbf{k_3}$ must form a closed triangle in Fourier space. Because of rotational invariance, instead, $B(k_1,k_2,k_3)$ depends only on the magnitude of the three wave vectors. The bispectrum $B(k_1,k_2,k_3)$ can further be rewritten as \cite{Babich:2004gb,Fergusson:2008ra,Akrami:2019izv}
\begin{equation}\label{eq:shape}
B(k_1,k_2,k_3)= f_{\text{NL}} \frac{S(k_1,k_2,k_3)}{(k_1k_2k_3)^2}\, ,
\end{equation}
where $f_{\text{NL}}$ is a dimensionless parameter quantifying the amplitude of the bispectrum\footnote{The exact definition of $f_{\rm NL}$ is fixed except for a constant normalization that may vary depending on the literature and the kind of primordial bispectrum under consideration. As an example, within the Planck mission it has been chosen to normalize the $f_{\text{NL}}$ coefficient of the scalar bispectrum such that \cite{Akrami:2019izv}
\begin{equation}
f^\zeta_{\text{NL}} =  \frac{5}{18} \, \frac{B_{\zeta}(k,k,k)}{P^2_\zeta(k)} \, .
\end{equation}
Here $B_{\zeta}(k_1,k_2,k_3)$ and $P_\zeta(k)$ denote respectively the scalar bispectrum and power spectrum from inflation.} and the shape function $S(k_1,k_2,k_3)$ encodes the functional dependence of the bispectrum on the specific triangle configurations. Typically, the shape function is normalized such that $S(k,k,k)=1$ in the equilateral limit, where the three momenta are equal. Notice that, due to the fact that the momenta form a closed triangle, once we specify two of the three momenta, the last one is automatically fixed.  As a consequence, the shape function depends only on the ratios between two of the three momenta and the third one (at least for almost scale invariant bispectra), e.g. $x_2 = k_2/k_1$ and $x_3 = k_3/k_1$.

The bispectrum of primordial tensor modes (evaluated at a given conformal time $\tau$) can be computed by means of the in-in formalism (see, e.g., \cite{Maldacena:2002vr,Weinberg:2005vy,Chen:2006nt,Chen:2010xka}) as
\begin{equation}
\label{eq:in-in-formula}
\langle\gamma_{s_1}(\mathbf{k_1},\tau)\gamma_{s_2}(\mathbf{k_2},\tau)\gamma_{s_3}(\mathbf{k_3},\tau)\rangle=-i\int_{-\infty}^{\tau}d\tau'\langle 0|\left[\gamma_{s_1}(\mathbf{k_1},\tau)\gamma_{s_2}(\mathbf{k_2},\tau)\gamma_{s_3}(\mathbf{k_3},\tau),H_{\text{int}}^{\gamma\gamma\gamma}(\tau')\right]|0\rangle\, ,
\end{equation}  
where $H_{\text{int}}^{\gamma\gamma\gamma}=-L_{\text{int}}^{\gamma\gamma\gamma}$ denotes the interaction Hamiltonian at cubic order in tensor perturbations.
In the computation of the in-in integrals we will adopt the usual $i\epsilon$-prescription, which amounts to deform the contour of integration by making the rotation $-\infty\rightarrow -\infty(1-i\epsilon)$ in the complex plane (see, e.g., \cite{Maldacena:2002vr}). The $i\epsilon$ contribution turns off the interactions in the far past and projects onto the vacuum state of the free theory. We are in particular interested in evaluating the primordial bispectrum on super-horizon scales, i.e. taking the limit $\tau\rightarrow 0$ of Eq. \eqref{eq:in-in-formula}.

\subsection{Graviton bispectra for constant coupling functions}

We start with the computation of the bispectrum from the Lagrangian PV1. The explicit expressions of the operators of $\mathcal{L}_{\text{PV1}}$ at cubic order in tensor perturbations, as well as the interaction Hamiltonian, can be found in App. \ref{App:Interaction-Hamiltonian}. We can then plug the interaction Hamiltonian \eqref{eq:PV1-Hamiltonian} into the in-in formula \eqref{eq:in-in-formula} and use the Wick theorem, with the contractions between the fields that are given by definition by
\begin{equation}
\langle 0|\gamma_{s_1}(\mathbf{k_1},0)\gamma_{s_2}(\mathbf{k},\tau)\epsilon^{(s_2)}_{ij}(\mathbf{k})|0\rangle=(2\pi)^3\delta_{s_1s_2}\delta^{(3)}(\mathbf{k_1}+\mathbf{k})u_{s_1}(\mathbf{k_1},0)u^*_{s_2}(\mathbf{k},\tau)\epsilon^{(s_2)*}_{ij}(\mathbf{k})\, ,
\end{equation} 
\begin{equation}
\langle 0|\gamma_{s_1}(\mathbf{k_1},0)\gamma'_{s_2}(\mathbf{k},\tau)\epsilon^{(s_2)}_{ij}(\mathbf{k})|0\rangle=(2\pi)^3\delta_{s_1s_2}\delta^{(3)}(\mathbf{k_1}+\mathbf{k})u_{s_1}(\mathbf{k_1},0)u^{*\prime}_{s_2}(\mathbf{k},\tau)\epsilon^{(s_2)*}_{ij}(\mathbf{k})\, .
\end{equation}
The bispectrum of primordial tensor perturbations can then be written as\footnote{The delta functions in the momenta lead to terms like $\epsilon^{(s)*}_{ij}(-\mathbf{k})$, but $\epsilon^{(s)*}_{ij}(-\mathbf{k})=\epsilon^{(s)}_{ij}(\mathbf{k})$ \eqref{eq:circ_identities}. The delta functions in the polarization indices instead reduce the initial nine polarizations (3+6 in the interaction Hamiltonian) down to only three.}
\begin{equation}
\label{eq:PV1-bispectrum}
\begin{split}
\langle\gamma_{s_1}(\mathbf{k_1})\gamma_{s_2}(\mathbf{k_2})\gamma_{s_3}(\mathbf{k_3})\rangle_{\text{PV1}}&=(2\pi)^3\delta^{(3)}(\mathbf{k_1}+\mathbf{k_2}+\mathbf{k_3})\,4\,\text{Im}\Big[I_1\,C_1^{s_1s_2s_3}(\mathbf{k_1},\mathbf{k_2},\mathbf{k_3})\\&
+I_2\,C_2^{s_1s_2s_3}(\mathbf{k_1},\mathbf{k_2},\mathbf{k_3})+I_3\,C_3^{s_1s_2s_3}(\mathbf{k_1},\mathbf{k_2},\mathbf{k_3})\\&
+I_4\,C_4^{s_1s_2s_3}(\mathbf{k_1},\mathbf{k_2},\mathbf{k_3})+I_5\,C_5^{s_1s_2s_3}(\mathbf{k_1},\mathbf{k_2},\mathbf{k_3})\\&
+I_6\,C_6^{s_1s_2s_3}(\mathbf{k_1},\mathbf{k_2},\mathbf{k_3})+I_7\,C_7^{s_1s_2s_3}(\mathbf{k_1},\mathbf{k_2},\mathbf{k_3})\\&
+I_8\,C_8^{s_1s_2s_3}(\mathbf{k_1},\mathbf{k_2},\mathbf{k_3})+I_9\,C_9^{s_1s_2s_3}(\mathbf{k_1},\mathbf{k_2},\mathbf{k_3})\Big]+\text{perm.}\,(k_i) \, ,
\end{split}
\end{equation}
where we have defined
\begin{align}
I_1&\equiv-u_{s_1}(k_1,0)u_{s_2}(k_2,0)u_{s_3}(k_3,0)\int_{-\infty}^{0} \frac{d\tau'}{\tau'}\,(f_1+g_1)\,\frac{du_{s_1}^*}{d\tau'}(k_1,\tau')\frac{du_{s_2}^*}{d\tau'}(k_2,\tau')u_{s_3}^*(k_3,\tau') \, ,  \nonumber
\\
I_2&\equiv-u_{s_1}(k_1,0)u_{s_2}(k_2,0)u_{s_3}(k_3,0)\int_{-\infty}^{0}d\tau'\,\frac{f_1}{2}\,\frac{du_{s_1}^*}{d\tau'}(k_1,\tau')\frac{du_{s_2}^*}{d\tau'}(k_2,\tau')\frac{du_{s_3}^*}{d\tau'}(k_3,\tau') \, ,  \nonumber
\\
I_3&\equiv\, u_{s_1}(k_1,0)u_{s_2}(k_2,0)u_{s_3}(k_3,0)\int_{-\infty}^{0}d\tau'\,\frac{(f_1+g_1)}{2}\,\frac{du_{s_1}^*}{d\tau'}(k_1,\tau')u_{s_2}^*(k_2,\tau')u_{s_3}^*(k_3,\tau') \, , \nonumber
\\
I_4&\equiv\, u_{s_1}(k_1,0)u_{s_2}(k_2,0)u_{s_3}(k_3,0)\int_{-\infty}^{0}d\tau'\,\frac{g_1}{2}\,u_{s_1}^*(k_1,\tau')u_{s_2}^*(k_2,\tau')\frac{du_{s_3}^*}{d\tau'}(k_3,\tau') \, , \nonumber
\\
I_5&\equiv-u_{s_1}(k_1,0)u_{s_2}(k_2,0)u_{s_3}(k_3,0)\int_{-\infty}^{0}d\tau'\,\left(f_1+\frac{g_1}{2}\right)\,\frac{du_{s_1}^*}{d\tau'}(k_1,\tau')u_{s_2}^*(k_2,\tau')u_{s_3}^*(k_3,\tau') \, , \nonumber
\\
I_6&\equiv\, u_{s_1}(k_1,0)u_{s_2}(k_2,0)u_{s_3}(k_3,0)\int_{-\infty}^{0}d\tau'\,\frac{f_1}{2}\,\frac{du_{s_1}^*}{d\tau'}(k_1,\tau')u_{s_2}^*(k_2,\tau')u_{s_3}^*(k_3,\tau') \, , \nonumber
\\
I_7&\equiv-u_{s_1}(k_1,0)u_{s_2}(k_2,0)u_{s_3}(k_3,0)\int_{-\infty}^{0} \frac{d\tau'}{\tau'}\,(f_1+g_1)\,u_{s_1}^*(k_1,\tau')\frac{du_{s_2}^*}{d\tau'}(k_2,\tau')\frac{du_{s_3}^*}{d\tau'}(k_3,\tau') \, , \nonumber
\\
I_8&\equiv\, u_{s_1}(k_1,0)u_{s_2}(k_2,0)u_{s_3}(k_3,0)\int_{-\infty}^{0} d\tau'\,\frac{(f_1+g_1)}{2}\,u_{s_1}^*(k_1,\tau')u_{s_2}^*(k_2,\tau')\frac{du_{s_3}^*}{d\tau'}(k_3,\tau') \, , \nonumber
\\
\label{eq:PV1-integral}
I_{9}&\equiv-u_{s_1}(k_1,0)u_{s_2}(k_2,0)u_{s_3}(k_3,0)\int_{-\infty}^{0}d\tau'\,\frac{g_1}{2}\,\frac{du_{s_1}^*}{d\tau'}(k_1,\tau')\frac{du_{s_2}^*}{d\tau'}(k_2,\tau')\frac{du_{s_3}^*}{d\tau'}(k_3,\tau')\, ,
\end{align}
where the $C_i^{s_1s_2s_3}(\mathbf{k_1},\mathbf{k_2},\mathbf{k_3})$, whose full expressions can be found in App. \ref{App:PV1-C_i}, are defined in terms of contractions between the wave vectors and the polarization tensors.

As a first approximation we assume that $a(\tau)\simeq -1/(H\tau)$, which holds at leading order in slow-roll, and take the Hubble parameter $H$ as constant during inflation. In the same spirit, we approximate the exact graviton mode function with the mode function in a de Sitter space-time. For tensor modes, this is given by (see, e.g., \cite{Maldacena:2002vr,Chen:2010xka})
\begin{equation}
\label{eq:de-Sitter-mode-function}
u_s(k,\tau)=\frac{iH}{M_{Pl}\sqrt{k^3}}\thinspace(1+ik\tau)\thinspace e^{-ik\tau} \, .	
\end{equation}	
This is justified because the corrections to the mode function \eqref{eq:de-Sitter-mode-function} that arise in this model are proportional to $\epsilon$ and $H/M_{\text{PV1}}$, which are both very small during inflation. So, using the de Sitter mode function gives the leading order contribution to the bispectrum. For the moment we also take the coupling functions $f_1$ and $g_1$ to be constant, leaving the more general case of time dependent couplings for the next section. \\
We can now solve analytically the integrals in Eq. \eqref{eq:PV1-integral}, which give
\begin{align}
&I_1=(f_1+g_1)\left(\frac{H^6}{M_{Pl}^6k_1^3k_2^3k_3^3}\right)k_1^2k_2^2\left[\frac{1}{k_T^2}+2\frac{k_3}{k_T^3}\right] \, , \\[5pt]&
I_2=\frac{f_1}{2}\left(\frac{H^6}{M_{Pl}^6k_1^3k_2^3k_3^3}\right)k_1^2k_2^2k_3^2\frac{6}{k_T^4} \, ,\\[5pt]&
I_3=\frac{f_1+g_1}{2}\left(\frac{H^6}{M_{Pl}^6k_1^3k_2^3k_3^3}\right)k_1^2\left[\frac{1}{k_T^2}+2\frac{k_2+k_3}{k_T^3}-6\frac{k_2k_3}{k_T^4}\right] \, , \\[5pt]&
I_4=\frac{g_1}{2}\left(\frac{H^6}{M_{Pl}^6k_1^3k_2^3k_3^3}\right)k_3^2\left[\frac{1}{k_T^2}+2\frac{k_1+k_2}{k_T^3}-6\frac{k_1k_2}{k_T^4}\right] \, , \\[5pt]&
I_5=-\left(f_1+\frac{g_1}{2}\right)\left(\frac{H^6}{M_{Pl}^6k_1^3k_2^3k_3^3}\right)k_1^2\left[\frac{1}{k_T^2}+2\frac{k_2+k_3}{k_T^3}-6\frac{k_2k_3}{k_T^4}\right] \, , \\[5pt]&
I_6=\frac{f_1}{2}\left(\frac{H^6}{M_{Pl}^6k_1^3k_2^3k_3^3}\right)k_1^2\left[\frac{1}{k_T^2}+2\frac{k_2+k_3}{k_T^3}-6\frac{k_2k_3}{k_T^4}\right] \, , \\[5pt]&
I_7=(f_1+g_1)\left(\frac{H^6}{M_{Pl}^6k_1^3k_2^3k_3^3}\right)k_2^2k_3^2\left[\frac{1}{k_T^2}+2\frac{k_1}{k_T^3}\right] \, , \\[5pt]&
I_8=\frac{f_1+g_1}{2}\left(\frac{H^6}{M_{Pl}^6k_1^3k_2^3k_3^3}\right)k_3^2\left[\frac{1}{k_T^2}+2\frac{k_1+k_2}{k_T^3}-6\frac{k_1k_2}{k_T^4}\right] \, , \\[5pt]&
I_9=\frac{g_1}{2}\left(\frac{H^6}{M_{Pl}^6k_1^3k_2^3k_3^3}\right)k_1^2k_2^2k_3^2\frac{6}{k_T^4} \, ,
\end{align} 	
where we have defined the total momentum $k_T\equiv k_1+k_2+k_3$. Notice that all the previous integrals give real results. Since all the $C_i^{s_1s_2s_3}(\mathbf{k_1},\mathbf{k_2},\mathbf{k_3})$ terms in Eq. \eqref{eq:PV1-bispectrum} are real (this is true also for the contributions proportional to $i\epsilon^{ijl}$ in Eqs. \eqref{eq:C4}-\eqref{eq:C5}-\eqref{eq:C7}-\eqref{eq:C8}, as can be easily checked by direct computations), the full bispectrum vanishes:
\begin{equation}
\langle\gamma_{s_1}(\mathbf{k_1})\gamma_{s_2}(\mathbf{k_2})\gamma_{s_3}(\mathbf{k_3})\rangle_{\text{PV1}}=0 \, .
\end{equation}
Hence, no parity-violating signatures arise in the graviton bispectrum if the couplings $f_1$ and $g_1$ are taken to be constant. We have found that our cubic interactions, even if not trivial, contribute to the non-linear graviton wave-function only via a pure (field-dependent) phase (see \cite{Maldacena:2011nz,Liu:2019fag,Bordin:2020eui} for more details on this regards), which does not affect super-horizon correlators. Notice that this is true under the approximation of using the de Sitter mode function for tensor modes. We expect that, using the exact solutions from the equation of motion of PGWs, may lead to a non-vanishing result, but with contributions that are suppressed by slow-roll parameters and the ratio $H/M_{\text{PV1}}$ (see \cite{Soda:2011am} for a detailed analysis of this issue in the case of slow-roll inflation with the parity violating Weyl cubic term). 

However, since there are no reasons for the coupling functions to be constant throughout all inflation, it is interesting to study the more general scenario in which the couplings are free to vary with time. In this case, using the de Sitter mode function will give the leading order non-vanishing contribution. 

Before doing this, let us make a similar analysis for the Lagrangian PV2. Once again, we refer the reader to App. \ref{App:Interaction-Hamiltonian} for the full expression of the interaction Hamiltonian at third order in tensor perturbations, which is reported in Eq. \eqref{eq:PV2-Hamiltonian}. By plugging this into the in-in formula \eqref{eq:in-in-formula}, we find
\begin{equation}
\label{eq:PV2-bispectrum}
\begin{split}
\langle\gamma_{s_1}(\mathbf{k_1})\gamma_{s_2}(\mathbf{k_2})\gamma_{s_3}(\mathbf{k_3})\rangle_{\text{PV2}}&=(2\pi)^3\delta^{(3)}(\mathbf{k_1}+\mathbf{k_2}+\mathbf{k_3})\,\text{Im}\bigg[\frac{\lambda_{s_1}}{2}\tilde{I}_1\,k_1\epsilon_{(s_1)}^{mi}(\mathbf{k_1})\epsilon^{(s_2)}_{mr}(\mathbf{k_2})\epsilon^r_{(s_3)i}(\mathbf{k_3})\\&
-\frac{i}{2}(\tilde{I}_2+\tilde{I}_3)\epsilon^{ijl}k_{1r}\epsilon_l^{(s_1)m}(\mathbf{k_1})\epsilon_j^{(s_2)r}(\mathbf{k_2})\epsilon^{(s_3)}_{mi}(\mathbf{k_3})\bigg]+\text{perm.}\,(k_i)\, ,
\end{split}
\end{equation}
where we have defined
\begin{align}
&\tilde{I}_1\equiv -u_{s_1}(k_1,0)u_{s_2}(k_2,0)u_{s_3}(k_3,0)\int_{-\infty}^{0}\frac{d\tau'}{\tau'}\left[\tilde{b}_1\frac{M_{Pl}}{H}-b\right] \frac{du_{s_1}^*}{d\tau'}(k_1,\tau')\frac{du_{s_2}^*}{d\tau'}(k_2,\tau')u_{s_3}^*(k_3,\tau')\, , \nonumber\\
&\tilde{I}_2\equiv -u_{s_1}(k_1,0)u_{s_2}(k_2,0)u_{s_3}(k_3,0)\int_{-\infty}^{0}\frac{d\tau'}{\tau'}\left[\tilde{b}_1\frac{M_{Pl}}{H}-b\right] u_{s_1}^*(k_1,\tau')\frac{du_{s_2}^*}{d\tau'}(k_2,\tau')\frac{du_{s_3}^*}{d\tau'}(k_3,\tau')\, , \nonumber \\
\label{eq:PV2-integral}
&\tilde{I}_3\equiv u_{s_1}(k_1,0)u_{s_2}(k_2,0)u_{s_3}(k_3,0)\int_{-\infty}^{0}d\tau'\,\frac{b}{2} \,\frac{du_{s_1}^*}{d\tau'}(k_1,\tau')\frac{du_{s_2}^*}{d\tau'}(k_2,\tau')\frac{du_{s_3}^*}{d\tau'}(k_3,\tau')\, .
\end{align}
Also in this case, we first work under the approximation of constant coupling functions and using de Sitter mode functions given in Eq. \eqref{eq:de-Sitter-mode-function} for gravitons, taking the Hubble parameter $H$ to be constant during inflation. The integrals in \eqref{eq:PV2-integral} can then be easily solved and give
\begin{align}
&\tilde{I}_1=\left[\tilde{b}_1\frac{M_{Pl}}{H}-b\right]\left(\frac{H^6}{M_{Pl}^6k_1^3k_2^3k_3^3}\right)k_1^2k_2^2\left[\frac{1}{k_T^2}+2\frac{k_3}{k_T^3}\right] \, ,\\[5pt]&
\tilde{I}_2=\left[\tilde{b}_1\frac{M_{Pl}}{H}-b\right]\left(\frac{H^6}{M_{Pl}^6k_1^3k_2^3k_3^3}\right)k_2^2k_3^2\left[\frac{1}{k_T^2}+2\frac{k_1}{k_T^3}\right] \, ,\\[5pt]&
\tilde{I}_3=-3b\left(\frac{H^6}{M_{Pl}^6k_1^3k_2^3k_3^3}\right)\frac{k_1^2k_2^2k_3^2}{k_T^4} \, .
\end{align}
Just like in the PV1 case, all the integrals give real contributions and thus none of the parity-breaking operators contributes to the bispectrum of primordial tensor modes:
\begin{equation}
\langle\gamma_{s_1}(\mathbf{k_1})\gamma_{s_2}(\mathbf{k_2})\gamma_{s_3}(\mathbf{k_3})\rangle_{\text{PV2}}=0\, .
\end{equation}
\subsection{Graviton bispectra for time dependent coupling functions}

So far, we have demonstrated that the parity-breaking operators introduced in \cite{Crisostomi:2017ugk} give no contributions to the primordial tensor bispectrum, assuming the de Sitter mode function for gravitons and in the case of constant coupling functions. However, as already discussed, there are no theoretical reasons why the couplings should be really constant during all the inflationary phase. Indeed, the couplings in Eqs. \eqref{f1g1}-\eqref{b1-tilde} depend on the scalar field and its kinetic term in full generality. Moreover, they could acquire a non-trivial time evolution because they might also depend on fields that are different from the inflaton field, being therefore not necessarily limited to a slow-roll evolution. 
Furthermore, it has been shown, e.g. in Ref. \cite{Shiraishi:2011st} for the parity-violating Weyl cubic terms, that, even if these operators do not contribute to the graviton bispectrum for constant couplings in the de Sitter limit \cite{Soda:2011am}, they can instead leave non-vanishing signatures if the couplings are free to vary with time.

Motivated by these reasons, we now extend the analysis of the previous section to the more general scenario where the coupling functions are allowed to evolve with time during inflation, and investigate whether parity-breaking signatures may arise in the primordial graviton bispectrum.  

We start by analyzing the PV1 model (defined by the Lagrangian \eqref{eq:PV2-action-ADM}). We restrict in particular to the interesting case where
\begin{equation}
\label{eq:PV1-constraint-no-ghost}
f_1+g_1=0 \, ,    
\end{equation}
such that the model is free from instabilities (see Eqs.~(\ref{PV1-A-B}),(\ref{MPV1}),
(\ref{f1g1}), and the discussion after
Eq.~(\ref{PV1-f-g})). Indeed, when the condition \eqref{eq:PV1-constraint-no-ghost} is realized, the cut-off scale $M_{\text{PV1}} \rightarrow \infty$ and all the corrections to the quadratic action for tensor modes \eqref{PV1-quadratic} disappear (this can also be seen directly from the equations of motion \eqref{PV1-eom-mode-functions}). In particular, this means that we recover $c_{T,s} = 1$ and $\chi = 0$.
Notice also that in such a case the theory makes sense without requiring an effective field theory treatment. If we renounce to \eqref{eq:PV1-constraint-no-ghost}, then the couplings have to obey $H/M_{PV1} \ll 1$, which will significantly limit also the amplitude of tensor bispectra. 

In order to make explicit computations, we need to assume a specific form of the coupling functions. In the rest of the paper we consider the case of a dilaton-like coupling, that naturally arises in theories with extra dimensions, like string theory. This can be written as $f_1=e^{(\phi-\phi_*)/M}$, where $M$ is some arbitrary energy scale. 
In slow-roll inflation this leads to a coupling that is simply given by a power of the conformal time\footnote{
Indeed, in slow-roll inflation the equation of motion for the inflaton field is
\begin{equation}
\label{eq:inflaton-eom}
\phi'\simeq \pm \sqrt{2\epsilon}M_{Pl}\tau^{-1} \, ,    
\end{equation}
where the $+$ and $-$ signs are for $\partial V/\partial \phi > 0$ and $\partial V/\partial \phi < 0$ respectively. This can be integrated to give
\begin{equation}
\phi=\phi_* \pm \sqrt{2\epsilon}M_{Pl} \ln\left(\frac{\tau}{\tau_*}\right) \, . 
\end{equation}
Substituting this into the exponential dilaton coupling, we end up with Eq. \eqref{eq:PV1-dilaton-couplings}.
} (see \cite{Shiraishi:2011st} for more details)
\begin{equation}
\label{eq:PV1-dilaton-couplings}
f_1(\tau)=\left(\frac{\tau}{\tau_*}\right)^A \, ,\qquad A=\pm\sqrt{2\epsilon}\,\frac{M_{Pl}}{M} \, ,
\end{equation} 
where $\epsilon$ is the usual slow-roll parameter, defined as in Eq. \eqref{eq:slow-roll-parameter}. The value of $\tau_*$ is fixed by the initial condition of Eq. \eqref{eq:inflaton-eom}. For example, we can take $\tau_*$ to be the time when the scale corresponding to present observable Universe crosses the horizon during inflation, such that $|\tau_*|=k_*^{-1}\sim 14$ Gpc \cite{Shiraishi:2011st}. In this case, $f_1$ would be of order unity for the current cosmological scales.

Thanks to the constraint \eqref{eq:PV1-constraint-no-ghost}, the integrals $I_1$, $I_3$, $I_7$, $I_8$ in Eq. \eqref{eq:PV1-integral} vanish,
\begin{equation}
\label{eq:PV1-integral-vanishing-dilaton}
I_1=I_3=I_7=I_8=0 \, .    
\end{equation}
As regarding the other integrals, once we have replaced the explicit expression of the mode function \eqref{eq:de-Sitter-mode-function} for gravitons in \eqref{eq:PV1-integral}, we end up with integrals of this kind
\begin{align}
 I_{(n,A)} = \tau_*^{-A} \int_{-\infty}^0 d\tau \, \tau^{n + A} e^{ i k_T \tau} \, ,
\end{align}
with $n+A>-1$. These can be analytically performed (with the usual $i \epsilon$ prescription) and give
\begin{align}
\label{eq:integrals-dilaton-couplings}
 I_{(n,A)} &= \tau_*^{-A}  (-1)^{(A + n)} \, ( i k_T)^{(-1 - A - n)} \, \Gamma(n + 1 + A) 
 \nonumber \\[5pt]
               &=  (-1)^{n} \,\, (-i)^{(- n - 1)}\, (n+A)! \,(- k_T \, \tau_*)^{-A} (- k_T)^{(- n - 1)}  \left[ \cos\left(\frac \pi 2 A\right) + i \sin \left(\frac \pi 2 A\right) \right] \, ,
\end{align}
where $\Gamma(x) = (x-1)!$ is the Gamma function and in the second equality we have used the Euler's formula. Using Eq. \eqref{eq:integrals-dilaton-couplings} we find
\begin{align}
\label{eq:PV1-integral-2-dilaton}
&I_2=-I_9=\frac{(3+A)!}{2}\left(\frac{H^6}{M_{Pl}^6k_1^3k_2^3k_3^3}\right)\frac{k_1^2k_2^2k_3^2}{k_T^4}\,(-k_T\tau_*)^{-A}\left[\cos\left(\frac{\pi}{2}A\right)+i\,\sin\left(\frac{\pi}{2}A\right)\right] \, ,\\[15pt]
\begin{split}
\label{eq:PV1-integral-4-dilaton}
&I_4=\frac{1}{2}\left(\frac{H^6}{M_{Pl}^6k_1^3k_2^3k_3^3}\right)(-k_T\tau_*)^{-A}\,k_3^2\left[-\frac{(1+A)!}{k_T^2}-(2+A)!\,\frac{k_1+k_2}{k_T^3}+(3+A)!\, \frac{k_1k_2}{k_T^4}\right]\\[5pt]
&\quad\times\left[\cos\left(\frac{\pi}{2}A\right)+i\,\sin\left(\frac{\pi}{2}A\right)\right] \, ,
\end{split}\\
\begin{split}
\label{eq:PV1-integral-5-dilaton}
&I_5=-I_6=\frac{1}{2}\left(\frac{H^6}{M_{Pl}^6k_1^3k_2^3k_3^3}\right)(-k_T\tau_*)^{-A}\,k_1^2\left[-\frac{(1+A)!}{k_T^2}-(2+A)!\,\frac{k_2+k_3}{k_T^3}+(3+A)!\, \frac{k_2k_3}{k_T^4}\right]\\[5pt]
&\qquad\qquad\times\left[\cos\left(\frac{\pi}{2}A\right)+i\,\sin\left(\frac{\pi}{2}A\right)\right] \, .
\end{split}
\end{align}
Notice that for $A=0$ we recover the result of the previous section, since the imaginary parts of the integrals \eqref{eq:PV1-integral-2-dilaton}-\eqref{eq:PV1-integral-5-dilaton} vanish and the bispectrum receives no contributions from the PV1 operators. When instead $A \ne 0$ (i.e. for time dependent couplings), the imaginary parts of these integrals switch on parity-breaking signatures in the primordial tensor bispectrum. This can be computed by plugging Eqs. \eqref{eq:PV1-integral-2-dilaton}-\eqref{eq:PV1-integral-5-dilaton} into Eq. \eqref{eq:PV1-bispectrum}. By doing so, we find
\begin{equation}
\label{eq:PV1-bispectrum-final}
\begin{split}
 \langle\gamma_{s_1}(\mathbf{k_1})\gamma_{s_2}(\mathbf{k_2})\gamma_{s_3}(\mathbf{k_3})\rangle_{\text{PV1}}&=(2\pi)^3\delta^{(3)}(\mathbf{k_1}+\mathbf{k_2}+\mathbf{k_3})\left(\frac{H}{M_{Pl}}\right)^6\left(\frac{\tau_*}{\bar{\tau}}\right)^{-A}\sin\left(\frac{\pi}{2}A\right)\\&   
 \times 2B_{s_1s_2s_3}^{\text{PV1}}(\mathbf{k_1},\mathbf{k_2},\mathbf{k_3})+\text{perm.}\,(k_i) \, ,
\end{split}    
\end{equation}
where we have defined 
\begin{equation}
\begin{split}
B_{s_1s_2s_3}^{\text{PV1}}(\mathbf{k_1},\mathbf{k_2},\mathbf{k_3})=\frac{1}{k_1^3k_2^3k_3^3}&\bigg\{(3+A)!\,\frac{k_1^2k_2^2k_3^2}{k_T^4}\,T_1^{s_1s_2s_3}(\mathbf{k_1},\mathbf{k_2},\mathbf{k_3})+k_3^2\left[-\frac{(1+A)!}{k_T^2}\right.\\&
\left.\;\;-(2+A)!\,\frac{k_1+k_2}{k_T^3}+(3+A)!\,\frac{k_1k_2}{k_T^4}\right]T_2^{s_1s_2s_3}(\mathbf{k_1},\mathbf{k_2},\mathbf{k_3})\\&
\;\;+k_1^2\left[-\frac{(1+A)!}{k_T^2}-(2+A)!\,\frac{k_2+k_3}{k_T^3}+(3+A)!\,\frac{k_2k_3}{k_T^4}\right]\\&
\;\;\times T_3^{s_1s_2s_3}(\mathbf{k_1},\mathbf{k_2},\mathbf{k_3})\bigg\} \, .      
\end{split}
\end{equation}
The $T_i^{s_1s_2s_3}(\mathbf{k_1},\mathbf{k_2},\mathbf{k_3})$ are again defined in terms of contractions between the wave vectors and the polarization tensors. Their expressions, which can be written as linear combinations of the $C_i^{s_1s_2s_3}(\mathbf{k_1},\mathbf{k_2},\mathbf{k_3})$, can be found in App. \ref{App:PV1-C_i}.

In Eq. \eqref{eq:PV1-bispectrum-final} we have evaluated the bispectrum at the horizon-crossing time of the total momentum $k_T$, $\bar{\tau}=-1/k_T$ . In fact, it is well known that, when performing in-in integrals, the main contributions arise around the horizon crossing of the overall momentum $k_T$ in the case of derivative interactions, as in the models under study. 
Notice that, because of the term $(\tau_*/\bar{\tau})^{-A}$ arising due to the time dependence of the coupling, the amplitude of the bispectrum is scale-dependent. In particular, for values of $A<0$ the amplitude increases going to small scales. For $A>0$, instead, the amplitude of the graviton bispectrum increases going to large scales. 

\begin{figure}
    \centering
    \includegraphics[scale=0.75]{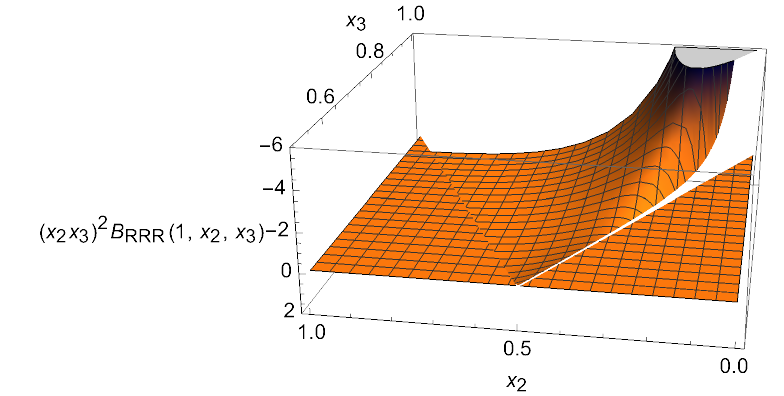}
    \includegraphics[scale=0.85]{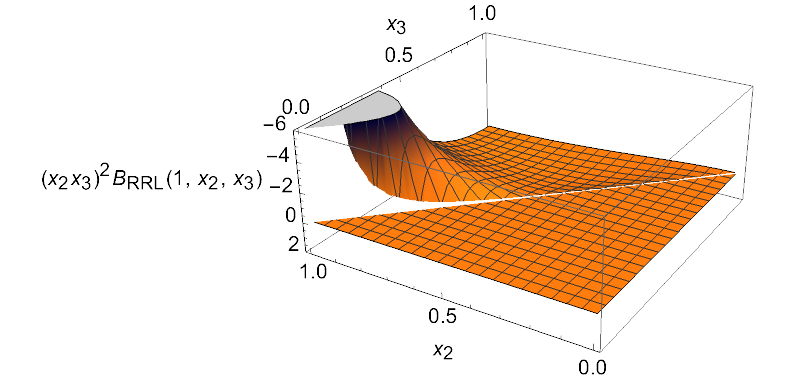}
    \caption{Shapes of the $RRR$ and $RRL$ graviton bispectra in the PV1 model, with $A=1$.
    The quantities $B_{RRR}(1,x_2,x_3)x_2^2x_3^3$ and $B_{RRL}(1,x_2,x_3)x_2^2x_3^3$ are plotted as functions of $x_2=k_2/k_1$ and $x_3=k_3/k_1$. They are both normalized to unity in the equilateral limit, $x_2=x_3=1$.}
    \label{fig:Shape-PV1}
\end{figure}

In Fig. \ref{fig:Shape-PV1} we plot the shape functions for two different polarization configurations, the first having $s_1=s_2=s_3=R$ and the second with $s_1=s_2=R$ and $s_3=L$. The other cases differ from these only by a minus sign, thus giving the same shape function. 
In particular, from the definition \eqref{eq:shape}, we show the quantities $B_{RRR}(1,x_2,x_3)x_2^2x_3^3$ and $B_{RRL}(1,x_2,x_3)x_2^2x_3^3$ as functions of $x_2=k_2/k_1$ and $x_3=k_3/k_1$. The plots are done assuming $A=1$, but we have checked that the qualitative behaviour of the shapes is independent from the value of $A$, which thus affects only the amplitude (and, in particular, its scale-dependence) of the primordial bispectra.
As already remarked in the previous sections, the shape function of the bispectrum is a powerful tool to discriminate among the various inflationary models. Indeed, different models of inflation contain different interaction terms between the dynamical fields of the theory, and thus leave distinctive signatures in the shapes of the primordial bispectra \cite{Akrami:2019izv}. 

To quantify how much a shape $S_1$ is similar to a reference shape $S_2$, it is common to introduce the cosine of the two shapes
\begin{equation} \label{eq:shape_cosine}
\cos(S_1,S_2)\equiv\frac{S_1\cdot S_2}{(S_1\cdot S_1)^{1/2}(S_2\cdot S_2)^{1/2}} \, ,    
\end{equation}
where the scalar product is defined as (see, e.g., \cite{Komatsu:2001rj,Babich:2004gb,Shiraishi:2011st})
\begin{equation}
S_1\cdot S_2\equiv \label{eq:shape_product} \sum_{k_i}\frac{S_1(k_1,k_2,k_3)S_2(k_1,k_2,k_3)}{P(k_1)P(k_2)P(k_3)} \, ,
\end{equation}
where the summation runs over all the wave vectors that form a triangle in the momentum space and $P(k)$ denotes the (tensor) power spectrum. By definition, the cosine is equal to 1 when $S_1=S_2$. Apart for an overall coefficient (that eventually cancels out when computing \eqref{eq:shape_cosine}), the quantity \eqref{eq:shape_product} can be expressed as an integral over $x_2$ and $x_3$ as\footnote{Notice that the ratio between two scales that enter in a given bispectrum can never be exactly $0$ or $1$, as in both cases we would deal with unphysical infinite-wavelength modes. This motivates the presence of the ratios $k_{\rm min}/k_{\rm max}$ in the extremes of integration in \eqref{eq:shape_product2}, so that the cosine between shapes depends on the ratio between the minimum and maximum scales that a given experiment can probe.}
\begin{equation} \label{eq:shape_product2}
S_1\cdot S_2 \propto \int_{\frac{k_{\rm min}}{k_{\rm max}}}^{1-\frac{k_{\rm min}}{k_{\rm max}}} dx_2 \int_{1-x_2}^{1-\frac{k_{\rm min}}{k_{\rm max}}} dx_3 \, x_2^4 x_3^4 \,  S_1(1,x_2,x_3) \, S_2(1,x_2,x_3)  \, .  
\end{equation}
As far as CMB experiments are concerned, in the following we evaluate the cosine between the shapes by summing over all the corresponding configurations in multipole space (using, as a first approximation, that $\ell \propto k$)  for multipoles $\ell$ ranging, in an ideal case, from $\ell_{\text{min}} = 2$ up to $\ell_{\text{max}}=1000$\footnote{In such a case, the choice of $\ell_{\text{max}}=1000$ is just indicative. Notice that this choice is rather optimistic if one considers bispectra involving the $B$-mode polarization field of the CMB (see, e.g.,  \cite{Meerburg:2016ecv,Duivenvoorden:2019ses}). However, we have explicitly checked that the cosines between shapes are not very sensitive to $\ell_{\text{max}}$ beyond $\ell_{\text{max}} \sim 100$.}. In the case of the shapes plotted in Fig. \ref{fig:Shape-PV1}, we find
\begin{equation}
\cos\left(B_{RRR}^{\text{PV1}},B_S\right)\simeq 
\begin{cases}
0.006 \, , \qquad S=\text{equilateral} \\
0.795 \, , \qquad S=\text{local} \\
0.328 \, , \qquad S=\text{orthogonal} 
\end{cases}
\end{equation}
for the $RRR$ case, while for the $RRL$ polarizations we have
\begin{equation}
\cos\left(B_{RRL}^{\text{PV1}},B_S\right)\simeq 
\begin{cases}
0.046 \, , \qquad S=\text{equilateral} \\
0.546 \, , \qquad S=\text{local} \\
0.299 \, , \qquad S=\text{orthogonal} \, . 
\end{cases}
\end{equation}
From these results and Fig. \ref{fig:Shape-PV1}, we realize that the maximum contributions come mainly from the squeezed configuration (corresponding to, e.g., $k_3 \ll k_1 \simeq k_2$).

We now compute the bispectrum from the PV2 Lagrangian (\ref{eq:PV2-action-ADM}), assuming time dependent couplings. Analogously to what we have done for the PV1 model, we restrict to the case with 
\begin{equation}\label{eq:PV2-constraint-no-ghost}
\tilde{b}_1M_{Pl}/H-b=0 \, ,    
\end{equation}
such that the model is free from instabilities. As a time dependent dilaton-like coupling we choose
\begin{equation}
\label{eq:PV2-dilaton-couplings}
b(\tau)=\left(\frac{\tau}{\tau_*}\right)^A \, ,\qquad A=\pm\sqrt{2\epsilon}\,\frac{M_{Pl}}{M} \, .
\end{equation} 
The integrals $\tilde{I}_1$ and $\tilde{I}_2$ in Eq. \eqref{eq:PV2-integral} vanish, while for $\tilde{I}_3$ we find
\begin{align}
\tilde{I}_3=-(3+A)!\left(\frac{H^6}{M_{Pl}^6k_1^3k_2^3k_3^3}\right)\frac{k_1^2k_2^2k_3^2}{k_T^4}\,(-k_T\tau_*)^{-A}\left[\text{cos}\left(\frac{\pi}{2}A\right)+i\,\text{sin}\left(\frac{\pi}{2}A\right)\right] \, .
\end{align} 
The bispectrum in Eq. \eqref{eq:PV2-bispectrum} can thus be rewritten as
\begin{equation}
\begin{split}
\langle\gamma_{s_1}(\mathbf{k_1})\gamma_{s_2}(\mathbf{k_2})\gamma_{s_3}(\mathbf{k_3})\rangle_{\text{PV2}}&=(2\pi)^3\delta^{(3)}(\mathbf{k_1}+\mathbf{k_2}+\mathbf{k_3})\,\frac{(3+A)!}{2}\left(\frac{H}{M_{Pl}}\right)^6\left(\frac{\tau_*}{\bar{\tau}}\right)^{-A}\sin\left(\frac{\pi}{2}A\right)\\&
\times B_{s_1s_2s_3}^{\text{PV2}}(\mathbf{k_1},\mathbf{k_2},\mathbf{k_3})+\text{perm.}\,(k_i) \, ,
\end{split}
\end{equation}
where we have defined
\begin{equation}
B_{s_1s_2s_3}^{\text{PV2}}(\mathbf{k_1},\mathbf{k_2},\mathbf{k_3})=\frac{1}{k_1^3k_2^3k_3^3}\bigg\{\frac{k_1^2k_2^2k_3^2}{k_T^4}\left[i\epsilon^{ijl}k_{1r}\,\epsilon_l^{(s_1)m}(\mathbf{k_1})\epsilon_j^{(s_2)r}(\mathbf{k_2})\epsilon^{(s_3)}_{mi}(\mathbf{k_3})\right]\bigg\} \, .    
\end{equation}
Thus, also the PV2 operators give a non-vanishing contribution to the three graviton bispectrum in the case of time-dependent couplings. Notice that in the limit where $A=0$ we recover the result of the previous section and the parity-breaking signatures are not present anymore.

\begin{figure}
    \centering
    \includegraphics[scale=0.8]{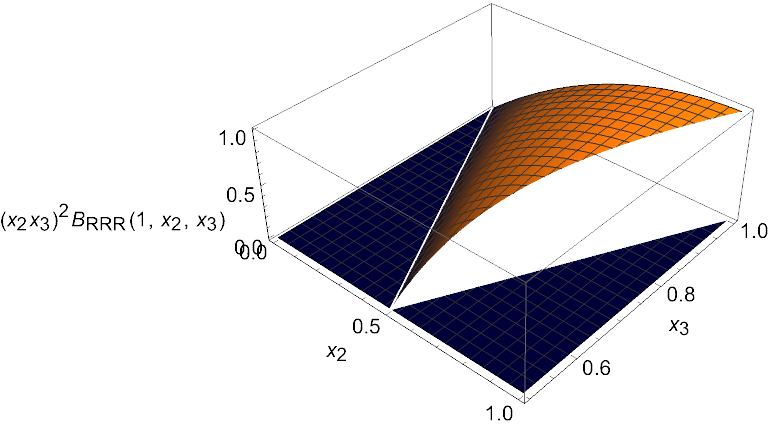}
    \includegraphics[scale=0.85]{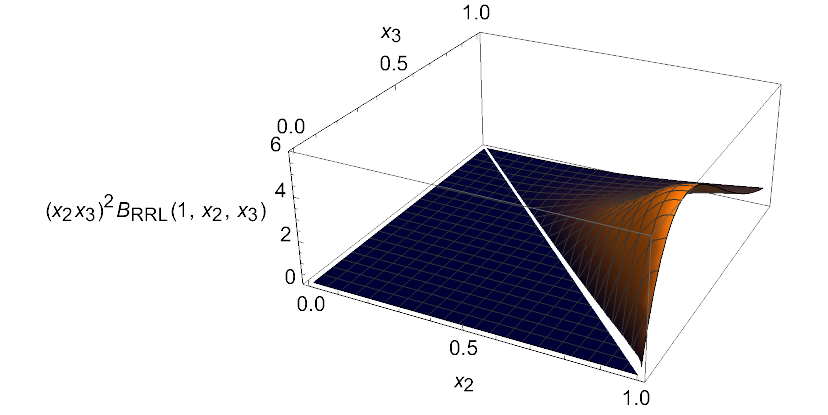}
    \caption{Shapes of the $RRR$ and $RRL$ graviton bispectra in the PV2 model, with $A=1$.
    The quantities $B_{RRR}(1,x_2,x_3)x_2^2x_3^3$ and $B_{RRL}(1,x_2,x_3)x_2^2x_3^3$ are plotted as functions of $x_2=k_2/k_1$ and $x_3=k_3/k_1$. They are both normalized to unity in the equilateral limit, $x_2=x_3=1$.}
    \label{fig:Shape-PV2}
\end{figure}

In Fig. \ref{fig:Shape-PV2} we plot the shape functions for the polarization configurations $RRR$ and $RRL$.
As in the previous case, we can compute the cosine of the shape functions. For $\ell_{\text{max}}=1000$, we find
\begin{equation}
\cos\left(B_{RRR}^{\text{PV2}},B_S\right)\simeq 
\begin{cases}
0.999 \, , \qquad S=\text{equilateral} \\
0.421 \, , \qquad S=\text{local} \\
0.205 \, , \qquad S=\text{orthogonal} 
\end{cases}
\end{equation}
for the $RRR$ case, and
\begin{equation}
\cos\left(B_{RRL}^{\text{PV2}},B_S\right)\simeq 
\begin{cases}
0.700 \, , \qquad S=\text{equilateral} \\
0.429 \, , \qquad S=\text{local} \\
0.095 \, , \qquad S=\text{orthogonal} 
\end{cases}
\end{equation}
for the $RRL$ polarizations. The $RRR$ shape peaks in the equilateral configuration (corresponding to $k_1 \simeq k_2 \simeq k_3$), as can also be seen directly from Fig. \ref{fig:Shape-PV2}. The $RRL$ shape receives instead a non-negligible contribution also in the squeezed configuration.

\subsection{Comments and observational prospects}

An interesting feature of the bispectra just derived is that they potentially lead to the breaking of the single field slow-roll consistency relation for tensor bispectra \cite{Maldacena:2002vr,Hinterbichler:2013dpa,Mirbabayi:2014zpa, Bordin:2016ruc}\footnote{It has been argued (see, e.g., \cite{Tanaka:2011aj,Pajer:2013ana,Creminelli:2013cga,Cabass:2016cgp}) that in the context of single field slow-roll models of inflation primordial bispectra in the squeezed limit (when one of the $k_i$ modes is much smaller than the others) correspond to a gauge artifact: in such a case one can perform a residual gauge transformation, passing from global coordinates to Conformal Fermi Coordinates (CFC), the latter being the coordinate frame of an observer that follows inflation in the background perturbed by this long-wavelength mode $k_L$. After performing this coordinate transformation, the squeezed limit bispectrum vanishes at leading order in the $k_L/k_S$ ratio, with corrections that, by the virtue of the equivalence principle, must be quadratic in $k_L/k_S$. However, recently in \cite{Matarrese:2020why} it has been shown that at least for the case of the scalar bispectrum this gauge artifact is valid only in the unphysical exactly infinite-wavelength limit where the long mode $k_L = 0$, while for physical modes $k_L \neq 0$ we lose this residual gauge freedom so that the consistency relations are indeed physical and measurable.}. These relations allow to predict the strict squeezed limit behaviour of the primordial bispectra in terms of the primordial power spectra \cite{Maldacena:2002vr,Creminelli:2004yq,Tanaka:2011aj,Creminelli:2011rh,Dai:2013kra,Creminelli:2013mca,Creminelli:2013cga,Pajer:2013ana,Hinterbichler:2013dpa,Mirbabayi:2014zpa,Bordin:2016ruc,Cabass:2016cgp}. According to these and neglecting the small scale dependence of the power spectra, the squeezed limit expression of the 3-graviton bispectrum is predicted to be
\begin{align} \label{eq:tensor_consistency}
\langle\gamma_{s_1}(\mathbf{k_1})\gamma_{s_2}(\mathbf{k_2})\gamma_{s_3}(\mathbf{k_3})\rangle|_{k_1\rightarrow 0} =  (2\pi)^3\delta^{(3)}(\mathbf{k_1}+\mathbf{k_2}+\mathbf{k_3})\, \frac{3}{2}\, P^{s_1}_T(k_1) P^{s_2}_T(k_2) \, \epsilon_{ij}^{(s_1)} \frac{k_2^i k_2^j}{k_2^2} \, \delta_{s_2 s_3} \, . 
\end{align}
 In the literature we already find some examples of models breaking the tensor consistency relation, such as scenarios where we have either additional particle content (see, e.g., \cite{Dimastrogiovanni:2015pla,Bordin:2016ruc,Iacconi:2020yxn}), violation of the adiabatic evolution of tensor perturbations (see, e.g., \cite{Kundu:2013gha,Bartolo:2013msa,Ozsoy:2019slf}), or violation of spatial diffeomorphisms (see, e.g., \cite{Bartolo:2015qvr,Ricciardone:2016lym}).

In our case, we have found that both $RRR$ ($LLL$) and $RRL$ ($LLR$) tensor bispectra originated by the new PV1 parity breaking operators give a nonzero contribution in the squeezed limit (see Fig. \ref{fig:Shape-PV1}), which is proportional to the couplings $f_1$ and $g_1$. On the contrary, primordial tensor power spectra, being the same of general relativity due to the constraint \eqref{eq:PV1-constraint-no-ghost}, do not depend by either $f_1$ or $g_1$. The result is that this squeezed signal is not predicted by Eq. \eqref{eq:tensor_consistency}, leading to the violation of the 3-graviton consistency relation. A similar pattern is also shared by the $RRL$ ($LLR$) bispectrum originated by the PV2 parity breaking operators\footnote{The reason why the $RRR$ ($LLL$) bispectrum originated by PV2 operators does not follow this behavior seems to be that, in the case where one correlates the same polarizations states, the squeezed contributions arise from the operators with more spatial derivatives, e.g. those in Eq. \eqref{eq:PV1-action-ADM} which contain the Riemann or Ricci tensors in the PV1 case. However, in the PV2 model these operators do not give any contribution to tensor perturbations (see App. \ref{App:Interaction-Hamiltonian}), as they also contain the derivative of the lapse function, which vanishes since $N=1$. Because of this, in the PV2 model we find a signal that peaks in the equilateral shape when the three polarizations are equal, but peaks in between the equilateral and squeezed limits for bispectra involving mixed polarization states. A similar feature has been found in the bispectra of PGWs in Horava-Lifshitz gravity for some of the operators, as noticed in \cite{Huang:2013epa}. We expect however that squeezed contributions may be present in the PV2 model in mixed correlators between scalar and tensor perturbations, since in this case also the operators with more spatial derivatives contribute to the final result.
}.
Notice that a similar pattern has been found previously in \cite{Huang:2013epa,Zhu:2013fja} in the context of studies of primordial non-Gaussianities in Horava-Lifshitz gravity: in such a case we get squeezed parity violating tensor bispectra that violate the tensor consistency relation, as they get a nonzero contribution in the exact squeezed limit that is not completely predicted by how primordial tensor power spectra in the right-hand side of \eqref{eq:tensor_consistency} are modified by Horava-Lifshitz gravity.

We argue that this analogy is not accidental, but related to the fact that both the theories introduce parity violating operators that violate also the Lorentz symmetry. In this regards, notice that this signature can be found only in particular late-time observables sensitive to the parity violation. In fact, for instance the modification introduced to the overall tensor bispectra (when summing over all the tensor polarizations) is equal to zero, due to the fact that the graviton bispectra under consideration have odd parity. Namely, they obey
\begin{align}
  \Braket{\gamma_{R}(\mathbf{k_1}) \gamma_{R}(\mathbf{k_2})  \gamma_{R}(\mathbf{k_3})}= - \Braket{\gamma_{L}(-\mathbf{k_1}) \gamma_{L}(-\mathbf{k_2})  \gamma_{L}(-\mathbf{k_3})} \, , \label{eq:parity_odd} 
\end{align}
and analogously for $RRL$ bispectrum versus $LLR$. 

Finally, let us discuss the observational prospects for detecting these parity-breaking signatures in the primordial bispectra through CMB experiments.
In order to find out CMB bispectra sensitive to our parity breaking signature, we follow the same reasoning as in \cite{Shiraishi:2011st,Shiraishi:2014ila,Bartolo:2018elp,Shiraishi:2019yux}. The starting point is the expression of the following spherical harmonic coefficients of the temperature ($X = T$) and E/B-mode polarization ($X = E/B$) anisotropies from the two circular polarizations of tensor perturbations \cite{Shiraishi:2010sm,Shiraishi:2010kd}
\begin{align} \label{eq:a_tensor}
  a_{\ell m}^{(t) X} =
4\pi (-i)^{\ell} \int \frac{d^3 k}{(2\pi)^{3}}
{\cal T}_{\ell(t)}^{X}(k)  \sum_{s = \pm 2} \left(\frac{s}{2}\right)^x \gamma_{s}(\mathbf{k}) \,  {}_{-s} Y_{\ell m}^*(\hat{k}) \, , 
\end{align}
where for convenience of the subsequent notation we have defined $\gamma_{\pm 2}(\mathbf{k}) \equiv \gamma_{R/L}(\mathbf{k})$. Here, ${}_{s} Y_{\ell m}(\hat{k})$ denotes a spin-weighted spherical harmonic, ${\cal T}_{\ell (t)}^{X}(k)$ is the tensor transfer function, and $x \equiv 0\, (1)$ for $X = T, E \, (B)$. Using \eqref{eq:a_tensor}, the CMB bispectra sourced by the primordial graviton bispectra can be written as 
\begin{align}
  \Braket{a_{\ell_1 m_1}^{(t) X_1} a_{\ell_2 m_2}^{(t) X_2} a_{\ell_3 m_3}^{(t) X_3} }
  =& \prod_{n=1}^3 4\pi (-i)^{\ell_n} \int \frac{d^3 k_n}{(2\pi)^{3}}
{\cal T}_{\ell_n (t)}^{X_n}(k_n)  \sum_{s_n = \pm 2} \left(\frac{s_n}{2}\right)^{x_n}  {}_{-s_n} Y_{\ell_n m_n}^*(\hat{k}_n)  \nonumber \\
&  \times \, \Braket{\gamma_{s_1}(\mathbf{k_1}) \gamma_{s_2}(\mathbf{k_2}) \gamma_{s_3}(\mathbf{k_3})}  \, .
 \label{eq:CMB_bis}
 \end{align}
Using the following well-known property of the weighted spherical harmonics 
\begin{align}
{}_{-s} Y_{\ell m}(-\hat{k}) = (-1)^\ell \, {}_{s} Y_{\ell m}(\hat{k}) \, ,
\end{align}
we can rewrite \eqref{eq:CMB_bis} as 
\begin{align}
  \Braket{a_{\ell_1 m_1}^{(t) X_1} a_{\ell_2 m_2}^{(t) X_2} a_{\ell_3 m_3}^{(t) X_3} }
  =& \prod_{n=1}^3 4\pi (-i)^{\ell_n} \int \frac{d^3 \vec{k}_n}{(2\pi)^{3}}
{\cal T}_{\ell_n (t)}^{X_n}(k_n)  \sum_{s_n = \pm 2} \left(\frac{s_n}{2}\right)^{x_n}  {}_{- s_n} Y_{\ell_n m_n}^*(\hat{k}_n) 
\nonumber \\ 
& \times \, (-1)^{x_1 + x_2 + x_3 + \ell_1+ \ell_2 + \ell_3 } \Braket{\gamma_{-s_1}(-\mathbf{k_1}) \gamma_{-s_2}(- \mathbf{k_2})\gamma_{-s_3}(- \mathbf{k_3})}\, . \label{eq:CMB_bis2}
 \end{align}
Now, matching Eq. \eqref{eq:CMB_bis} with \eqref{eq:CMB_bis2} under the parity-odd condition \eqref{eq:parity_odd}, we find that
\begin{equation}
  \Braket{a_{\ell_1 m_1}^{(t) X_1} a_{\ell_2 m_2}^{(t) X_2} a_{\ell_3 m_3}^{(t) X_3} }[1 + (-1)^{x_1 + x_2 + x_3 + \ell_1+ \ell_2 + \ell_3 }] = 0
\end{equation}
must always hold, independently by the kind of CMB modes that we are cross-correlating. Thus, a non-vanishing contribution to the CMB angular bispectrum is confined to the following multipole configurations
\begin{equation}
x_1 + x_2 + x_3+ \ell_1+ \ell_2 + \ell_3 = {\rm odd} \, . \label{eq:odd_condition}
\end{equation}
It is worth stressing that these combinations are not realized by the usual parity-conserving theories like Einstein gravity (for which the sum defined in \eqref{eq:odd_condition} has to be even) and therefore they can become robust indicators of parity breaking models with odd tensor bispectra if they are detected.

According to the latest forecasts and previsions (see, e.g., \cite{Meerburg:2016ecv,Shiraishi:2019yux,DeLuca:2019jzc,Duivenvoorden:2019ses}), forthcoming CMB experiments focusing on the polarization field (like, e.g., the LiteBIRD mission) will be able to probe order 1 amplitudes of tensor squeezed non-Gaussianities through the measurement of the CMB angular bispectra involving the $B$ modes. This would justify a more detailed analysis of the detection prospects of these models in CMB angular bispectra in view of next experiments focusing on the search for the $B$-mode polarization field of the CMB.

\section{Conclusions}
\label{conclusions}

In this work, we have firstly reviewed the parity breaking signatures in primordial tensor power spectrum arising from chiral scalar-tensor theories with higher order derivatives proposed in \cite{Crisostomi:2017ugk}. We have shown that, due to the possible appearance of instabilities of one of the two circular polarizations, the final amount of chirality in PGWs is theoretically constrained to be very small in a way that it is going to be challenging to measure with current and forthcoming experiments. We have also remarked that the final prediction on the level of chirality \eqref{PV1-chi} and \eqref{PV2-chi} is expected to be degenerate with that of other parity violating operators previously studied in the context of inflationary models. 

Thus, with the idea of breaking this degeneracy, we have made an original analysis of the parity breaking effects of these operators on the graviton bispectrum statistics. We have shown that, taking de Sitter mode functions and assuming constant couplings, the graviton self-interactions contribute to the non-linear graviton wave-function only via a pure phase, thus not affecting the graviton bispectrum.

Therefore, we have fixed a particular setup where the effects of the operators under study are vanishing on the tensor power spectrum, removing the need for an effective field theory treatment of the theory. We thus have assumed time dependent couplings and computed the corresponding signature in the graviton bispectra for all the possible kind of polarization combinations. In such a case, we got non-vanishing tensor bispectra which peak in the squeezed and equilateral configurations (see Figs. \ref{fig:Shape-PV1} and \ref{fig:Shape-PV2}), and are also characterized by odd parity \eqref{eq:parity_odd}. As also previously shown, both these peculiarities can improve the Signal-to-Noise ratio detection of primordial bispectra in CMB experiments focusing on the $B$-mode detection, making our case study relevant for the forthcoming CMB experiments aiming to improve the measurements of the CMB polarization fields (like, e.g., the LiteBIRD satellite). This would motivate a subsequent forecast on the detection prospects of the models under consideration using CMB bispectra. 

We have finally noticed that the computed bispectra, written in the circular basis, lead to the breaking of tensor consistency relations. We argue that this is a manifestation of the simultaneous violation of parity and Lorentz symmetries, leaving a more general study in this direction for future works.

\paragraph{Acknowledgements}

We thank Matteo Biagetti, Giovanni Cabass, Alexander Ganz, Sabino Matarrese and Gianmassimo Tasinato for useful comments and valuable discussions on this work. G.O. acknowledges partial financial support by ASI Grant No. 2016-24-H.0. N.B. and L.C. acknowledge support from the COSMOS network (www.cosmosnet.it) through the ASI (Italian Space Agency) Grants 2016-24-H.0 and 2016-24-H.1-2018. 
\newpage
\appendix

\section{Interaction Hamiltonians at cubic order in tensor perturbations}
\label{App:Interaction-Hamiltonian}

\subsection{PV1 interaction Hamiltonian}

Assuming $N=1$ and $N_i=0$ for the reason explained in Sec. \ref{CST theories}, the PV1 Lagrangian in the ADM formalism under the constraints \eqref{eq:constraint_a} can be written as \cite{Crisostomi:2017ugk}
\begin{align}
\label{eq:PV1-action-ADM}
\nonumber
\sqrt{-g}\,\mathcal{L}_{\text{PV1}}=\frac{2\dot{\phi}^2}{M_{Pl}^4}\epsilon^{ijl}&\left[2(2a_1+a_2+4a_4)\left(KK_{mi}D_lK_j^{\thinspace m}+\prescript{(3)}{\noindent}{R_{mi}}D_lK_j^{\thinspace m}-K_{mi}K^{mn}D_lK_{jn}\right)\right.\\&
\left.-(a_2+4a_4)\left(2K_{mi}K_j^{\thinspace n}D_nK_l^{\thinspace m}+\prescript{(3)}{\noindent}{R_{jlm}}^nD_nK_i^{\thinspace m}\right)\right] \, ,
\end{align}
where $K_{ij}=\dot{h}_{ij}/2$ is the extrinsic curvature tensor, $K=h^{ij}K_{ij}$ its trace, $D_i$ denotes the three-dimensional covariant derivative, $\prescript{(3)}{\noindent}{R_{mi}}$ and $\prescript{(3)}{\noindent}{R_{jlm}}^n$ are the three-dimensional Ricci and Riemann tensors respectively. Notice that we have reintroduced the Planck mass by dimensional analysis.
By expanding $\gamma_{ij}$ up to third order, the operators present in \eqref{eq:PV1-action-ADM} at cubic order in tensor perturbations take the following forms
\begin{align}
\epsilon^{ijl}KK_{mi}D_lK_j^{\thinspace m}&=\frac{3}{4}\dot{a}a\,\epsilon^{ijl}\left[(\partial_l\dot{\gamma}_j^{\thinspace m})\dot{\gamma}_{mr}\gamma^r_{\thickspace i}+(\partial_r\gamma_l^{\thinspace m})\dot{\gamma}_j^{\thinspace r}\dot{\gamma}_{mi}\right],
\\
\nonumber 
\epsilon^{ijl}\prescript{(3)}{\noindent}{R_{mi}}D_lK_j^{\thinspace m}&=\frac{1}{4}\epsilon^{ijl}\left[-\frac{1}{2}(\partial_r\partial^r\gamma_{mi})(\partial_l\gamma_{jk})\dot{\gamma}^{km}-\frac{1}{2}(\partial_r\partial^r\gamma_{mi})\gamma_{jk}(\partial_l\dot{\gamma}^{km})\right.\\& \nonumber
\left.\qquad\quad\;\;\;-\frac{1}{2}(\partial_r\partial^r\gamma_{mi})(\partial_k\gamma_l^{\thinspace m})\dot{\gamma}_j^{\thickspace k}+\frac{1}{2}(\partial_r\partial^r\gamma_{mi})(\partial^m\gamma_{lk})\dot{\gamma}_j^{\thinspace k}\right.\\& \nonumber
\left.\qquad\quad\;\;\;+(\partial_l\dot{\gamma}_j^{\thinspace m})\gamma^{kr}(\partial_k\partial_r\gamma_{mi})+(\partial_l\dot{\gamma}_j^{\thinspace m})(\partial_m\gamma^{kr})(\partial_k\gamma_{ir})\right.\\&  \nonumber
\left.\qquad\quad\;\;\;-(\partial_l\dot{\gamma}_j^{\thinspace m})\gamma^{kr}(\partial_k\partial_i\gamma_{mr})-\frac{1}{2}(\partial_l\dot{\gamma}_j^{\thinspace m})\gamma^r_{\thickspace i}(\partial_k\partial^k\gamma_{mr})\right.\\& 
\left.\qquad\quad\;\;\;-\frac{1}{2}(\partial_l\dot{\gamma}_j^{\thinspace m})(\partial_i\gamma_r^{\thinspace k})(\partial_m\gamma_k^{\thickspace r})-(\partial_l\dot{\gamma}_j^{\thinspace m})(\partial^k\gamma_{ri})(\partial^r\gamma_{km})\right],
\\
\epsilon^{ijl}K_{mi}K^{mn}D_lK_{jn}&=\epsilon^{ijl}\left[-\frac{1}{2}\dot{a}a\,(\partial_r\gamma_l^{\thinspace m})\dot{\gamma}_{jm}\dot{\gamma}_i^{\thinspace r}+\frac{1}{2}\dot{a}a(\partial_l\dot{\gamma}_{jr})\gamma_{mi}\dot{\gamma}^{mr}+\frac{1}{8}a^2(\partial_l\dot{\gamma}_{jr})\dot{\gamma}_{mi}\dot{\gamma}^{mr}\right],
\\
\epsilon^{ijl}K_{mi}K_j^{\thinspace n}D_nK_l^{\thinspace m}&=\frac{1}{4}\epsilon^{ijl}\left[\dot{a}a\,(\partial_j\dot{\gamma}_l^{\thinspace m})\dot{\gamma}_{mr}\gamma^r_{\thickspace i}+\dot{a}a\,(\partial_r\gamma_j^{\thinspace m})\dot{\gamma}_l^{\thinspace r}\dot{\gamma}_{mi}+\frac{1}{2}a^2(\partial_r\dot{\gamma}_l^{\thinspace m})\dot{\gamma}_j^{\thinspace r}\dot{\gamma}_{mi}\right],
\\
\nonumber
\epsilon^{ijl}\prescript{(3)}{\noindent}{R_{jlm}}^nD_nK_i^{\thinspace m}&=\frac{1}{4}\epsilon^{ijl}\bigg[-\gamma_{lr}(\partial_m\partial_j\gamma^{rn})(\partial_n\dot{\gamma}_i^{\thinspace m})+\gamma_{lr}(\partial^n\partial_j\gamma^{r}_{\thickspace m})(\partial_n\dot{\gamma}_i^{\thinspace m})\\& \nonumber
\qquad\quad\;\;\;\thinspace-2\gamma^{pn}(\partial_m\partial_l\gamma_{pj})(\partial_n\dot{\gamma}_i^{\thinspace m})-2\gamma^{pn}(\partial_p\partial_j\gamma_{lm})(\partial_n\dot{\gamma}_i^{\thinspace m})\\& \nonumber
\qquad\quad\;\;\;\thinspace+(\partial_m\gamma_j^{\thickspace r})(\partial_r\gamma_l^{\thickspace n})(\partial_n\dot{\gamma}_i^{\thinspace m})-(\partial_m\gamma_j^{\thickspace r})(\partial^n\gamma_{lr})(\partial_n\dot{\gamma}_i^{\thinspace m})\\& \nonumber
\qquad\quad\;\;\;\thinspace
-(\partial^n\gamma_j^{\thickspace r})(\partial_r\gamma_{lm})(\partial_n\dot{\gamma}_i^{\thinspace m})-(\partial^n\gamma_{mr})(\partial_l\gamma^r_{\thickspace j})(\partial_n\dot{\gamma}_i^{\thinspace m})\\& \nonumber
\qquad\quad\;\;\;\thinspace+(\partial_r\gamma_{mj})(\partial_l\gamma^{nr})(\partial_n\dot{\gamma}_i^{\thinspace m})-(\partial_r\gamma_{mj})(\partial^r\gamma_l^{\thickspace n})(\partial_n\dot{\gamma}_i^{\thinspace m})\\& \nonumber
\qquad\quad\;\;\;\thinspace-(\partial_j\gamma_{rm})(\partial_l\gamma^{nr})(\partial_n\dot{\gamma}_i^{\thinspace m})+(\partial_j\gamma_{rm})(\partial^r\gamma_l^{\thickspace n})(\partial_n\dot{\gamma}_i^{\thinspace m})\\& \nonumber
\qquad\quad\;\;\;\thinspace
+2(\partial_m\partial_l\gamma^n_{\thickspace\, j})\dot{\gamma}_{ik}(\partial_n\gamma^{km})+(\partial_m\partial_l\gamma^n_{\thickspace\, j})\gamma_{ik}(\partial_n\dot{\gamma}^{km})\\& \nonumber
\qquad\quad\;\;\;\thinspace +(\partial^n\partial_j\gamma_{lm})\gamma_{ik}(\partial_n\dot{\gamma}^{km})-(\partial_m\partial_l\gamma^n_{\thickspace\, j})(\partial_i\gamma_n^{\thickspace k})\dot{\gamma}_k^{\thickspace m}\\& \nonumber
\qquad\quad\;\;\;\thinspace-(\partial^n\partial_j\gamma_{lm})(\partial_i\gamma_n^{\thickspace k})\dot{\gamma}_k^{\thickspace m}+(\partial_m\partial_l\gamma^n_{\thickspace\, j})(\partial^k\gamma_{ni})\dot{\gamma}_k^{\thickspace m}\\&
\qquad\quad\;\;\;\thinspace
+(\partial^n\partial_j\gamma_{lm})(\partial^k\gamma_{ni})\dot{\gamma}_k^{\thickspace m}-(\partial_m\partial_l\gamma^n_{\thickspace\, j})(\partial^m\gamma_{nk})\dot{\gamma}_i^{\thinspace k}\bigg].
\end{align}
From these, we can then compute the interaction Hamiltonian. In Fourier space, this reads
\begin{equation}
\label{eq:PV1-Hamiltonian}
\begin{split}
H_{\gamma\gamma\gamma}^{\text{PV1}}(\tau)&=\sum_{s_1,s_2,s_3}\int d^3k\int d^3p \int d^3q\ \frac{\delta^{(3)}(\mathbf{k}+\mathbf{p}+\mathbf{q})}{(2\pi)^6}2\Bigg\{\lambda_{s_1}(f_1+g_1)aHk\gamma^{\prime s_1}_{\mathbf{k}}\gamma^{\prime s_2}_{\mathbf{p}}\gamma^{s_3}_{\mathbf{q}}\\&
\times\epsilon_{(s_1)}^{mi}(\mathbf{k})\epsilon_{(s_2)}^{mr}(\mathbf{p})\epsilon_{(s_3)i}^r(\mathbf{q})
+\lambda_{s_2}\frac{(f_1+g_1)}{2}k^2p\gamma^{\prime s_1}_{\mathbf{k}}\gamma^{s_2}_{\mathbf{p}}\gamma^{s_3}_{\mathbf{q}}\epsilon^{(s_1)}_{mi}(\mathbf{k})\epsilon_k^{(s_2)i}(\mathbf{p})\epsilon_{(s_3)}^{km}(\mathbf{q})\\&
+\lambda_{s_3}(f_1+g_1)qk_kq_r\gamma^{\prime s_1}_{\mathbf{k}}\gamma^{s_2}_{\mathbf{p}}\gamma^{s_3}_{\mathbf{q}}\epsilon_j^{(s_1)m}(\mathbf{k})\epsilon_{(s_2)}^{kr}(\mathbf{p})\epsilon_m^{(s_3)j}(\mathbf{q})-\lambda_{s_1}\left(f_1+\frac{g_1}{2}\right)kp_mq_k\\&
\times\gamma^{\prime s_1}_{\mathbf{k}}\gamma^{s_2}_{\mathbf{p}}\gamma^{s_3}_{\mathbf{q}}\epsilon_{(s_1)}^{mi}(\mathbf{k})\epsilon_{(s_2)}^{kr}(\mathbf{p})\epsilon^{s_3}_{ir}(\mathbf{q})+\lambda_{s_1}\frac{f_1}{2}kq^2\gamma^{\prime s_1}_{\mathbf{k}}\gamma^{s_2}_{\mathbf{p}}\gamma^{s_3}_{\mathbf{q}}\epsilon_{(s_1)}^{mi}(\mathbf{k})\epsilon^r_{(s_2)i}(\mathbf{p})\epsilon^{(s_3)}_{mr}(\mathbf{q})\\&
+\lambda_{s_1}\frac{f_1}{2}kp_iq_m\gamma^{\prime s_1}_{\mathbf{k}}\gamma^{s_2}_{\mathbf{p}}\gamma^{s_3}_{\mathbf{q}}\epsilon_{(s_1)}^{mi}(\mathbf{k})\epsilon_r^{(s_2)k}(\mathbf{p})\epsilon_k^{(s_3)r}(\mathbf{q})+\lambda_{s_2}(f_1+g_1)pp^kq^r\gamma^{\prime s_1}_{\mathbf{k}}\gamma^{s_2}_{\mathbf{p}}\gamma^{s_3}_{\mathbf{q}}\\&
\times\epsilon_{(s_1)}^{mi}(\mathbf{k})\epsilon_r^{(s_2)j}(\mathbf{p})\epsilon^{(s_3)}_{km}(\mathbf{q})-\lambda_{s_1}\frac{f_1}{2}k\gamma^{\prime s_1}_{\mathbf{k}}\gamma^{\prime s_2}_{\mathbf{p}}\gamma^{\prime s_3}_{\mathbf{q}}\epsilon_r^{(s_1)i}(\mathbf{k})\epsilon_{mi}^{(s_2)}(\mathbf{p})\epsilon_{(s_3)}^{mr}(\mathbf{q})\\&
+\lambda_{s_1}\frac{g_1}{2}kk_np_m\gamma^{s_1}_{\mathbf{k}}\gamma^{s_2}_{\mathbf{p}}\gamma^{\prime s_3}_{\mathbf{q}}\epsilon_r^{(s_1)i}(\mathbf{k})\epsilon_{(s_2)}^{rn}(\mathbf{p})\epsilon_i^{(s_3)m}(\mathbf{q})+\lambda_{s_2}\frac{g_1}{2}k^2p\gamma^{s_1}_{\mathbf{k}}\gamma^{s_2}_{\mathbf{p}}\gamma^{\prime s_3}_{\mathbf{q}}\\&
\times\epsilon^{(s_1)}_{mr}(\mathbf{k})\epsilon_{(s_2)}^{ri}(\mathbf{p})\epsilon_i^{(s_3)m}(\mathbf{q})+\lambda_{s_1}\frac{g_1}{2}kk_mq_n\gamma^{s_1}_{\mathbf{k}}\gamma^{s_2}_{\mathbf{p}}\gamma^{\prime s_3}_{\mathbf{q}}\epsilon_{(s_1)}^{ni}(\mathbf{k})\epsilon^{(s_2)}_{ik}(\mathbf{p})\epsilon_{(s_3)}^{km}(\mathbf{q})\\&
+\lambda_{s_1}\frac{g_1}{2}k(\mathbf{k}\cdot\mathbf{p})\gamma^{s_1}_{\mathbf{k}}\gamma^{s_2}_{\mathbf{p}}\gamma^{\prime s_3}_{\mathbf{q}}\epsilon_m^{(s_1)i}(\mathbf{k})\epsilon^{(s_2)}_{ik}(\mathbf{p})\epsilon_{(s_3)}^{km})(\mathbf{q})
-\lambda_{s_1}\frac{g_1}{2}kk_mp_i\gamma^{s_1}_{\mathbf{k}}\gamma^{s_2}_{\mathbf{p}}\gamma^{\prime s_3}_{\mathbf{q}}\\&
\times\epsilon_{(s_1)}^{ni}(\mathbf{k})\epsilon_n^{(s_2)k}(\mathbf{p})\epsilon_k^{(s_3)m}(\mathbf{q})+\lambda_{s_1}\frac{g_1}{2}kk^np_i\gamma^{s_1}_{\mathbf{k}}\gamma^{s_2}_{\mathbf{p}}\gamma^{\prime s_3}_{\mathbf{q}}\epsilon_m^{(s_1)i}(\mathbf{k})\epsilon_n^{(s_2)k}(\mathbf{p})\epsilon_k^{(s_3)m}(\mathbf{q})\\&
+\lambda_{s_1}\frac{g_1}{2}kk_mp^k\gamma^{s_1}_{\mathbf{k}}\gamma^{s_2}_{\mathbf{p}}\gamma^{\prime s_3}_{\mathbf{q}}\epsilon_{(s_1)}^{ni}(\mathbf{k})\epsilon^{(s_2)}_{ni}(\mathbf{p})\epsilon_k^{(s_3)m}(\mathbf{q})-\lambda_{s_1}\frac{g_1}{2}kk^np^k\gamma^{s_1}_{\mathbf{k}}\gamma^{s_2}_{\mathbf{p}}\gamma^{\prime s_3}_{\mathbf{q}}\\&
\times\epsilon_m^{(s_1)i}(\mathbf{k})\epsilon^{(s_2)}_{ni}(\mathbf{p})\epsilon_k^{(s_3)m}(\mathbf{q})
+\epsilon^{ijl}\bigg[-i(f_1+g_1)aHk_r\gamma^{s_1}_{\mathbf{k}}\gamma^{\prime s_2}_{\mathbf{p}}\gamma^{\prime s_3}_{\mathbf{q}}\\&
\times\epsilon_l^{(s_1)m}(\mathbf{k})\epsilon_j^{(s_2)r}(\mathbf{p})\epsilon^{(s_3)}_{mi}(\mathbf{q})-i\frac{(f_1+g_1)}{2}k^2q_l\gamma^{s_1}_{\mathbf{k}}\gamma^{s_2}_{\mathbf{p}}\gamma^{\prime s_3}_{\mathbf{q}}\epsilon^{(s_1)}_{mi}(\mathbf{k})\epsilon^{(s_2)}_{jk}(\mathbf{p})\epsilon_{(s_3)}^{km}(\mathbf{q})\\&
-i\frac{(f_1+g_1)}{2}k^2p_k\gamma^{s_1}_{\mathbf{k}}\gamma^{s_2}_{\mathbf{p}}\gamma^{\prime s_3}_{\mathbf{q}}\epsilon^{(s_1)}_{mi}(\mathbf{k})\epsilon_l^{(s_2)m}(\mathbf{p})\epsilon_j^{(s_3)k}(\mathbf{q})+i\frac{(f_1+g_1)}{2}k^2p^m\gamma^{s_1}_{\mathbf{k}}\gamma^{s_2}_{\mathbf{p}}\gamma^{\prime s_3}_{\mathbf{q}}\\&
\times\epsilon^{(s_1)}_{mi}(\mathbf{k})\epsilon^{(s_2)}_{lk}(\mathbf{p})\epsilon_j^{(s_3)k}(\mathbf{q})+i\left(f_1+\frac{g_1}{2}\right)k_kp_lq_r\gamma^{\prime s_1}_{\mathbf{k}}\gamma^{s_2}_{\mathbf{p}}\gamma^{s_3}_{\mathbf{q}}\epsilon_j^{(s_1)m}(\mathbf{k})\epsilon_{(s_2)}^{kr}(\mathbf{p})\epsilon^{(s_3)}_{mi}(\mathbf{q})\\&
-i\left(f_1+\frac{g_1}{2}\right)k_np_lq_i\gamma^{\prime s_1}_{\mathbf{k}}\gamma^{s_2}_{\mathbf{p}}\gamma^{s_3}_{\mathbf{q}}\epsilon_j^{(s_1)m}(\mathbf{k})\epsilon_{(s_2)}^{nr}(\mathbf{p})\epsilon^{(s_3)}_{mr}(\mathbf{q})+i\left(f_1+\frac{g_1}{2}\right)p^kq_lq^r.\\&
\times\gamma^{\prime s_1}_{\mathbf{k}}\gamma^{s_2}_{\mathbf{p}}\gamma^{s_3}_{\mathbf{q}}\epsilon_j^{(s_1)m}(\mathbf{k})\epsilon_{ri}^{(s_2)}(\mathbf{p})\epsilon^{(s_3)}_{km}(\mathbf{q})+i\frac{g_1}{2}k_r\gamma^{\prime s_1}_{\mathbf{k}}\gamma^{\prime s_2}_{\mathbf{p}}\gamma^{\prime s_3}_{\mathbf{q}}\epsilon_l^{(s_1)m}(\mathbf{k})\epsilon_j^{(s_2)r}(\mathbf{p})\epsilon^{(s_3)}_{mi}(\mathbf{q})\\&
-i\frac{g_1}{2}(\mathbf{p}\cdot\mathbf{q})p_j\gamma^{s_1}_{\mathbf{k}}\gamma^{s_2}_{\mathbf{p}}\gamma^{\prime s_3}_{\mathbf{q}}\epsilon^{(s_1)}_{lr}(\mathbf{k})\epsilon_{(s_2)m}^r(\mathbf{p})\epsilon_i^{(s_3)m}(\mathbf{q})-i\frac{g_1}{2}k_mp_rq_n\gamma^{s_1}_{\mathbf{k}}\gamma^{s_2}_{\mathbf{p}}\gamma^{\prime s_3}_{\mathbf{q}}\\&
\times\epsilon_j^{(s_1)r}(\mathbf{k})\epsilon_l^{(s_2)n}(\mathbf{p})\epsilon_i^{(s_3)m}(\mathbf{q})-i\frac{g_1}{2}(\mathbf{k}\cdot\mathbf{p})k_m\gamma^{s_1}_{\mathbf{k}}\gamma^{s_2}_{\mathbf{p}}\gamma^{\prime s_3}_{\mathbf{q}}\epsilon_j^{(s_1)r}(\mathbf{k})\epsilon_{lr}^{(s_2)}(\mathbf{p})
\epsilon_i^{(s_3)m}(\mathbf{q})\bigg]\Bigg\},
\end{split}
\end{equation}	
where the prime denotes a derivative with respect to conformal time.

\subsection{PV2 interaction Hamiltonian}

Assuming $N=1$ and $N_i=0$, the PV2 Lagrangian in the ADM formalism under the constraints \eqref{eq:constraint_b} can be written as \cite{Crisostomi:2017ugk}
\begin{equation}
\label{eq:PV2-action-ADM}
\sqrt{-g}\,\mathcal{L}_{\text{PV2}}=2\frac{\dot{\phi}^3}{M_{Pl}^5}\epsilon^{ijl}\left[b_1K_{mi}D_lK_j^{\thinspace m}+\frac{(b_4+b_5-b_3)}{M_{Pl}^3}\dot{\phi}K_{mi}K_j^{\thinspace n}D_nK_l^{\thinspace m}\right] \, .
\end{equation}
Notice that the operators in the second line of the action (3.23) in \cite{Crisostomi:2017ugk} do not give any contribution to tensor perturbations, since $D_iN=0$. 
At cubic order in tensor perturbations, the operators in \eqref{eq:PV2-action-ADM} read
\begin{align}
\epsilon^{ijl}K_{mi}D_lK_j^{\thinspace m}&=\frac{1}{4}a^2\,\epsilon^{ijl}\left[(\partial_l\dot{\gamma}_j^{\thinspace m})\dot{\gamma}_{mr}\gamma^r_{\thickspace i}+(\partial_r\gamma_l^{\thinspace m})\dot{\gamma}_j^{\thinspace r}\dot{\gamma}_{mi}\right],
\\
\epsilon^{ijl}K_{mi}K_j^{\thinspace n}D_nK_l^{\thinspace m}&=\frac{1}{4}\epsilon^{ijl}\left[\dot{a}a\,(\partial_j\dot{\gamma}_l^{\thinspace m})\dot{\gamma}_{mr}\gamma^r_{\thickspace i}+\dot{a}a\,(\partial_r\gamma_j^{\thinspace m})\dot{\gamma}_l^{\thinspace r}\dot{\gamma}_{mi}+\frac{1}{2}a^2(\partial_r\dot{\gamma}_l^{\thinspace m})\dot{\gamma}_j^{\thinspace r}\dot{\gamma}_{mi}\right].
\end{align}
The interaction Hamiltonian in Fourier space is thus given by
\begin{equation}
\label{eq:PV2-Hamiltonian}
\begin{split}
H_{\gamma\gamma\gamma}^{\text{PV2}}=&\sum_{s_1,s_2,s_3}\int d^3k\int d^3p \int d^3q\ \frac{\delta^{(3)}(\mathbf{k}+\mathbf{p}+\mathbf{q})}{(2\pi)^6}\left\{\frac{1}{2}\left[\tilde{b}_1M_{Pl}-bH\right]\Big[\lambda_{s_1}ak\gamma^{\prime s_1}_{\mathbf{k}}\gamma^{\prime s_2}_{\mathbf{p}}\gamma^{s_3}_{\mathbf{q}}\right.\\&
\left.\times\epsilon_{(s_1)}^{mi}(\mathbf{k})\epsilon_{mr}^{(s_2)}(\mathbf{p})\epsilon^r_{(s_3)i}(\mathbf{q})+i\epsilon^{ijl}ak_r\gamma^{s_1}_{\mathbf{k}}\gamma^{\prime s_2}_{\mathbf{p}}\gamma^{\prime s_3}_{\mathbf{q}}\epsilon_l^{(s_1)m}(\mathbf{k})\epsilon_j^{(s_2)r}(\mathbf{p})\epsilon_{mi}^{(s_3)}(\mathbf{q})\Big]\right.\\&
\left.+i\frac{b}{4}\epsilon^{ijl}k_r\gamma^{\prime s_1}_{\mathbf{k}}\gamma^{\prime s_2}_{\mathbf{p}}\gamma^{\prime s_3}_{\mathbf{q}}\epsilon_l^{(s_1)m}(\mathbf{k})\epsilon_j^{(s_2)r}(\mathbf{p})\epsilon_{mi}^{(s_3)}(\mathbf{q})\right\}.
\end{split}
\end{equation}

\section{Explicit expressions of $C_i^{s_1s_2s_3}(\mathbf{k_1},\mathbf{k_2},\mathbf{k_3})$ and $T_i^{s_1s_2s_3}(\mathbf{k_1},\mathbf{k_2},\mathbf{k_3})$}
\label{App:PV1-C_i}
We report here the complete expressions of the contributions that appear in the PV1 bispectrum \eqref{eq:PV1-bispectrum}: 
\begin{align}
\label{eq:C1}
&C_1^{s_1s_2s_3}(\mathbf{k_1},\mathbf{k_2},\mathbf{k_3})=C_2^{s_1s_2s_3}(\mathbf{k_1},\mathbf{k_2},\mathbf{k_3})=\lambda_{s_1}k_1\epsilon_{(s_1)}^{mi}(\mathbf{k_1})\epsilon^{(s_2)}_{mr}(\mathbf{k_2})\epsilon_{(s_3)i}^r(\mathbf{k_3}) \, , 
\\[10pt] & \nonumber
\label{eq:C3}
C_3^{s_1s_2s_3}(\mathbf{k_1},\mathbf{k_2},\mathbf{k_3})=\lambda_{s_2}k_2\left[k_1^2\epsilon_{mi}^{(s_1)}(\mathbf{k_1})\epsilon_k^{(s_2)i}(\mathbf{k_2})\epsilon_{(s_3)}^{km}(\mathbf{k_3})+2k_2^kk_3^r\,\epsilon_j^{(s_1)m}(\mathbf{k_1})\epsilon_r^{(s_2)j}(\mathbf{k_2})\epsilon_{km}^{(s_3)}(\mathbf{k_3})\right]
\\ & \qquad\qquad\qquad\qquad\,
+2\lambda_{s_3}k_3k_{1k}k_{3r}\,\epsilon_j^{(s_1)m}(\mathbf{k_1})\epsilon_{(s_2)}^{kr}(\mathbf{k_2})\epsilon_m^{(s_3)j}(\mathbf{k_3}) \, , 
\\[10pt] & \nonumber
\label{eq:C4}
C_4^{s_1s_2s_3}(\mathbf{k_1},\mathbf{k_2},\mathbf{k_3})=\left[\lambda_{s_2}k_1^2k_2+\lambda_{s_1}(\mathbf{k_1}\cdot\mathbf{k_2})k_1\right]\epsilon_{ri}^{(s_1)}(\mathbf{k_1})\epsilon_{(s_2)}^{im}(\mathbf{k_2})\epsilon_m^{(s_3)r}(\mathbf{k_3})\\& \nonumber
\qquad\qquad\qquad\qquad+\lambda_{s_1}k_1k_{1n}k_{2m}\Big[\epsilon_r^{(s_1)i}(\mathbf{k_1})\epsilon_{(s_2)}^{rn}(\mathbf{k_2})\epsilon_i^{(s_3)m}(\mathbf{k_3})-\epsilon_{(s_1)}^{rm}(\mathbf{k_1})\epsilon_r^{(s_2)i}(\mathbf{k_2})\epsilon_i^{(s_3)n}(\mathbf{k_3})\\& \nonumber
\qquad\qquad\qquad\qquad\qquad\qquad\qquad\;\;+\epsilon_r^{(s_1)m}(\mathbf{k_1})\epsilon_{(s_2)}^{ni}(\mathbf{k_2})\epsilon_i^{(s_3)r}(\mathbf{k_3})+\epsilon_{(s_1)}^{ri}(\mathbf{k_1})\epsilon_{ri}^{(s_2)}(\mathbf{k_2})\epsilon_{(s_3)}^{mn}(\mathbf{k_3})\\& \nonumber
\qquad\qquad\qquad\qquad\qquad\qquad\qquad\;\;-\epsilon_r^{(s_1)i}(\mathbf{k_1})\epsilon^{n}_{(s_2)i}(\mathbf{k_2})\epsilon_{(s_3)}^{mr}(\mathbf{k_3})\Big]\\& \nonumber
\qquad\qquad\qquad\qquad+\lambda_{s_1}k_1k_{1n}k_{3m}\,\epsilon_{(s_1)}^{mi}(\mathbf{k_1})\epsilon^{(s_2)}_{ir}(\mathbf{k_2})\epsilon_{(s_3)}^{rn}(\mathbf{k_3})\\& \nonumber
\qquad\qquad\qquad\qquad+i\epsilon^{ijl}\Big[(\mathbf{k_2}\cdot\mathbf{k_3})k_{2j}\,\epsilon^{(s_1)}_{lr}(\mathbf{k_1})\epsilon_{(s_2)m}^r(\mathbf{k_2})\epsilon_i^{(s_3)m}(\mathbf{k_3})\\& \nonumber
\qquad\qquad\qquad\qquad\qquad\quad  +k_{1m}k_{2r}k_{3n}\,\epsilon_j^{(s_1)r}(\mathbf{k_1})\epsilon_l^{(s_2)n}(\mathbf{k_2})\epsilon_i^{(s_3)m}(\mathbf{k_3})\\& 
\qquad\qquad\qquad\qquad\qquad\quad+(\mathbf{k_1}\cdot\mathbf{k_2})k_{1m}\,\epsilon_j^{(s_1)r}(\mathbf{k_1})\epsilon^{(s_2)}_{lr}(\mathbf{k_2})\epsilon_i^{(s_3)m}(\mathbf{k_3})\Big] \, ,
\\[10pt] & \nonumber
\label{eq:C5}
C_5^{s_1s_2s_3}(\mathbf{k_1},\mathbf{k_2},\mathbf{k_3})=\lambda_{s_1}k_1k_{2m}k_{3k}\,\epsilon_{(s_1)}^{mi}(\mathbf{k_1})\epsilon_{(s_2)}^{kr}(\mathbf{k_2})\epsilon_{ir}^{(s_3)}(\mathbf{k_3})\\&
\nonumber \qquad\qquad\qquad\qquad\,+i\epsilon^{ijl}\Big[k_{1k}k_{2l}k_{3r}\,\epsilon_j^{(s_1)m}(\mathbf{k_1})\epsilon_{(s_2)}^{kr}(\mathbf{k_2})\epsilon_{mi}^{(s_3)}(\mathbf{k_3})
\\ & \qquad\qquad\qquad\qquad\qquad\quad\, \nonumber
-k_{1k}k_{2l}k_{3i}\epsilon_j^{(s_1)m}(\mathbf{k_1})\epsilon_{(s_2)}^{kr}(\mathbf{k_2})\epsilon^{(s_3)}_{mr}(\mathbf{k_3})
\\ & \qquad\qquad\qquad\qquad\qquad\quad\,
+k_2^kk_{3l}k_3^r\,\epsilon_j^{(s_1)m}(\mathbf{k_1})\epsilon^{(s_2)}_{ri}(\mathbf{k_2})\epsilon^{(s_3)}_{km}(\mathbf{k_3})\Big] \, , 
\\[10pt] & 
\label{eq:C6}
C_6^{s_1s_2s_3}(\mathbf{k_1},\mathbf{k_2},\mathbf{k_3})=\lambda_{s_1}k_1\left[k_{2i}k_{3m}\,\epsilon_{(s_1)}^{mi}(\mathbf{k_1})\epsilon_r^{(s_2)k}(\mathbf{k_2})\epsilon_k^{(s_3)r}(\mathbf{k_3})+k_3^2\epsilon_{(s_1)}^{mi}(\mathbf{k_1})\epsilon_{(s_2)i}^r(\mathbf{k_2})\epsilon_{mr}^{(s_3)}(\mathbf{k_3})\right] \, , 
\\ & 
\label{eq:C7}
C_7^{s_1s_2s_3}(\mathbf{k_1},\mathbf{k_2},\mathbf{k_3})=C_9^{s_1s_2s_3}(\mathbf{k_1},\mathbf{k_2},\mathbf{k_3})=i\epsilon^{ijl}k_{1r}\,\epsilon_l^{(s_1)m}(\mathbf{k_1})\epsilon_j^{(s_2)r}(\mathbf{k_2})\epsilon_{mi}^{(s_3)}(\mathbf{k_3}) \, , 
\\[10pt] & \nonumber
\label{eq:C8}
C_8^{s_1s_2s_3}(\mathbf{k_1},\mathbf{k_2},\mathbf{k_3})=ik_1^2\epsilon^{ijl}\left[k_{3l}\,\epsilon^{(s_1)}_{mi}(\mathbf{k_1})\epsilon^{(s_2)}_{jk}(\mathbf{k_2})\epsilon_{(s_3)}^{km}(\mathbf{k_3})+k_{2k}\,\epsilon^{(s_1)}_{mi}(\mathbf{k_1})\epsilon_l^{(s_2)m}(\mathbf{k_2})\epsilon_j^{(s_3)k}(\mathbf{k_3})\right.
\\ & \qquad\qquad\qquad\qquad\qquad\qquad\,
\left.-k_2^m\epsilon^{(s_1)}_{mi}(\mathbf{k_1})\epsilon^{(s_2)}_{lk}(\mathbf{k_2})\epsilon_j^{(s_3)k}(\mathbf{k_3})\right] \, .
\end{align}

The $T_i^{s_1s_2s_3}(\mathbf{k_1},\mathbf{k_2},\mathbf{k_3})$ that enter in the final expression of the PV1 bispectrum \eqref{eq:PV1-bispectrum-final} with time-dependent couplings can be written in terms of the $C_i^{s_1s_2s_3}(\mathbf{k_1},\mathbf{k_2},\mathbf{k_3})$ as 
\begin{align}
&T_1^{s_1s_2s_3}(\mathbf{k_1},\mathbf{k_2},\mathbf{k_3})=C_2^{s_1s_2s_3}(\mathbf{k_1},\mathbf{k_2},\mathbf{k_3})-C_7^{s_1s_2s_3}(\mathbf{k_1},\mathbf{k_2},\mathbf{k_3})\, , \\
&T_2^{s_1s_2s_3}(\mathbf{k_1},\mathbf{k_2},\mathbf{k_3})=C_4^{s_1s_2s_3}(\mathbf{k_1},\mathbf{k_2},\mathbf{k_3})\, , \\
&T_3^{s_1s_2s_3}(\mathbf{k_1},\mathbf{k_2},\mathbf{k_3})=C_5^{s_1s_2s_3}(\mathbf{k_1},\mathbf{k_2},\mathbf{k_3})-C_6^{s_1s_2s_3}(\mathbf{k_1},\mathbf{k_2},\mathbf{k_3})\, .
\end{align}

\section{Polarization tensors}
In this section we set our conventions for the polarization tensors by fixing an explicit representation for them. We can first use the momentum conservation, $\mathbf{k_1}+\mathbf{k_1}+\mathbf{k_3}=0$, and the invariance under rotations to make the three wave vectors lying on the same plane, that we choose to be the $(x,y)$ plane:
\begin{equation}
\mathbf{k_1}=k_1(1,0,0)\, ,\quad \mathbf{k_2}=k_2(\cos\theta,\sin\theta,0)\, ,\quad \mathbf{k_3}=k_3(\cos\varphi,\sin\varphi,0)\, ,
\end{equation}
where $\theta$ and $\varphi$ are the angles that $\mathbf{k_1}$ forms with $\mathbf{k_2}$ and $\mathbf{k_3}$ respectively. Without loss of generality, we can choose $0\le\theta\le\pi$ and $\pi\le\varphi\le 2\pi$, such that
\begin{equation}
\cos\theta=\frac{k_3^2-k_1^2-k_2^2}{2k_1k_2}\, ,\quad \sin\theta=\frac{\lambda}{2k_1k_2}\, ,\quad \cos\varphi=\frac{k_2^2-k_3^2-k_1^2}{2k_3k_1}\, ,\quad \sin\varphi=-\frac{\lambda}{2k_3k_1}\, ,    
\end{equation}
with
\begin{equation}
\lambda=\sqrt{2k_1^2k_2^2+2k_2^2k_3^2+2k_3^2k_1^2-k_1^4-k_2^4-k_3^4}\, .    
\end{equation}
With this representation of the wave vectors we can then write explicitly the polarization tensors as \cite{McFadden:2011kk,Soda:2011am}
\begin{equation}
\epsilon^{(s)}(\mathbf{k_1})=\frac{1}{\sqrt{2}}
\begin{pmatrix}
0 & 0 & 0\\
0 & 1 & i\lambda_s\\
0 & i\lambda_s & -1\\
\end{pmatrix},
\end{equation}
\begin{equation}
\epsilon^{(s)}(\mathbf{k_2})=\frac{1}{\sqrt{2}}
\begin{pmatrix}
\sin^2\theta & -\sin\theta\cos\theta & -i\lambda_s\sin\theta\\
-\sin\theta\cos\theta & \cos^2\theta & i\lambda_s\cos\theta\\
-i\lambda_s\sin\theta & i\lambda_s\cos\theta & -1\\
\end{pmatrix},
\end{equation}
\begin{equation}
\epsilon^{(s)}(\mathbf{k_3})=\frac{1}{\sqrt{2}}
\begin{pmatrix}
\sin^2\varphi & -\sin\varphi\cos\varphi & -i\lambda_s\sin\varphi\\
-\sin\varphi\cos\varphi & \cos^2\varphi & i\lambda_s\cos\varphi\\
-i\lambda_s\sin\varphi & i\lambda_s\cos\varphi & -1\\
\end{pmatrix},
\end{equation}
where $\lambda_s=\pm 1$ for $s=R$ and $s=L$ respectively.

\bibliography{references.bib}

\end{document}